\documentclass[intlimits,twoside,a4paper]{article}
\usepackage[cp1251]{inputenc}
\usepackage[eqsecnum]{cmpj3}

\usepackage{bm}

\issue{2022}{25}{3}{33202}
\doinumber{10.5488/CMP.25.33202}

\title[Water-DMSO liquid mixtures]
{Revisiting the composition dependence of the properties of
water-dimethyl sulfoxide liquid mixtures.  
Molecular dynamics computer simulations}

\author[M. Aguilar, H. Dominguez, O. Pizio]
{M. Aguilar\orcid{0000-0003-3850-1188}\refaddr{label1}, 
H. Dominguez\orcid{0000-0001-6126-9300}\refaddr{label2},
        O. Pizio\orcid{0000-0001-8333-4652}\refaddr{label1}
\thanks{Corresponding author: \email{oapizio@gmail.com}.}}

\addresses{
\addr{label1}Instituto de Qu\'{i}mica, Universidad Nacional Aut\'{o}noma de M\'{e}xico,
Circuito Exterior, 04510, Cd. Mx., M\'{e}xico
\addr{label2} Instituto de Investigaciones en Materiales, Universidad Nacional Aut\'{o}noma de M\'{e}xico,
Circuito Exterior, 04510, Cd. Mx., M\'{e}xico}

\Keywords{molecular dynamics, water-dimethyl sulfoxide mixtures, surface tension, dielectric constant, partial
molar volumes, partial molar enthalpies}

\date{Received July 06, 2022, in final form August 09, 2022}

\begin{document}

\maketitle

\begin{abstract}
We have revisited the composition dependence of principal
properties of liquid water-DMSO mixtures
by using the isobaric-isothermal molecular dynamics computer simulations. A set of 
non-polarizable semi-flexible models for the DMSO molecule 
combined with the TIP4P-2005 and TIP4P/$\varepsilon$  water 
models is considered. We restrict calculations to
atmospheric pressure, 0.1013 MPa, and room temperature, 298.15~K.
Composition trends of density, excess mixing volume and excess mixing enthalpy,
partial molar volumes and partial molar enthalpies of species,
apparent molar volumes are reported. Besides, we explore composition trends of
the self-diffusion of species, the static dielectric constant and the surface tension.
Evolution of the microscopic structure of the mixture with composition is analyzed
in terms of radial distributions functions, coordination numbers and
the fractions of hydrogen-bonded molecules.
We intend to capture the peculiarities of mixing the species in the mixture 
upon the DMSO molar fraction and  the anomalous behaviors, 
if manifested in each of the properties under study. 
The quality of several combinations of the models for species is evaluated in detail
to establish the possibility of necessary improvements.

\printkeywords
%
%
\end{abstract}

\section{Introduction}

Liquid mixtures of dimethyl sulfoxide (DMSO) and water are of much practical importance as solvents
and reaction media in organic chemistry and chemical engineering~\cite{jacobs,martin1,martin2}.
Besides, they are widely used in several biological applications, cryoprotection being just
one example, see e.g.,~\cite{mcgann,murthy,rabin}. 

Thus, it is not only desirable but also necessary to
have an accurate description of various properties of these mixtures dependent on composition, $X$,
temperature, $T$, and pressure, $P$. Common experimental practice is to study the composition effects 
at a room temperature and not far from it. Moreover, in many studies the atmospheric pressure is assumed.
 Much less studies are focused on temperature and pressure trends of 
the behavior of water-DMSO liquid mixtures.
Unfortunately, experimental results do not cover all the 
cases of interest. In other words they do not provide comprehensive knowledge about the dependence
of even some basic properties of the system in question, upon changes of these 
thermodynamic variables. 
As documented in reference~\cite{wiewior}, the experimental knowledge and understanding 
of the microscopic structure and dynamic properties of the systems in question 
mainly follows from the application of  neutron scattering,
nuclear magnetic resonance, dielectric relaxation and vibrational spectroscopy methods.
In order to interpret the experimental observations in detail 
and to get ampler insights, one is
forced to resort to computer simulations methodology.

Usual strategy of computer simulations methodology is to choose a model of each species, 
water and DMSO in the present case,
and assume the cross interactions by using the combination rules. 
Then, a software based on strict rules of statistical mechanics is applied.
Adequacy of the computer simulations predictions for a given model for
solution, upon changing $X, T$ and $P$, is then tested by comparison with 
some reference experimental data.  
Afterwards, the profit of the method
is in the possibility to interpret these observables 
at microscopic level with molecular level details. 
Besides, it becomes possible to explore
a wider set of properties, in addition to the benchmark or reference experimental data, 
with a reasonable degree of confidence.
It is worth to mention that the trajectories of particles generated by molecular dynamics (MD)
simulations provide two types of results. Namely, one of them includes ``direct''
properties, e.g., the equation of state in terms of density at a given temperature and 
pressure upon changing composition in the NPT ensemble, enthalpy, $H(T,P,X)$,
a set of radial distribution functions of particles and  the self-diffusion coefficients of species.
On the other hand, another type of properties follows from the time averages of
fluctuations. These are, for example, the isothermal compressibility and the dielectric constant.
The latter are more sensitive to the number of particles in the principal
simulated volume, to the time extension of the simulations and to the method of account
for the long-range contribution of inter-particle interactions.

Having this in mind, the principal objective of the present work is to revisit 
the composition trends of the behavior of water-DMSO mixtures at atmospheric pressure 
and room temperature by using isobaric-isothermal molecular dynamics computer simulation. 
The reason is that several previous publications were focused on elucidation of
anomalies of composition behavior, or better say of non-monotonous behaviors of different 
properties and their interpretation from molecular dynamics simulations.
Some sets of experimental data permit to obtain anomalies straightforwardly. On the other
hand, hidden anomalies from apparently monotonous behavior can be extracted as well.
We would like to show that confidence into the presence of  anomalies 
coming from computer simulations results requires a very good description of the
underlying properties with respect to experimental data. Otherwise, the anomalies
resulting from the analyses of computer simulations results may follow
from the behavior of a model rather than from reality.

The first MD simulations of systems of
our interest were reported in~\cite{rao1,vaisman,luzar3,gunsteren1}. In order to formulate the
force fields, the experimental observations and quantum chemical calculations were used.
Further refinement of parameters was performed in some cases by minimizing the
differences of model calculations
with a few experimental observables. Concerning the nomenclature and parameters of the DMSO models
most frequently used in the literature, we refer to the Table I in each of the
references~\cite{luzar3,bako1,jedlovszky1}, as well as to the table 2 of reference~\cite{samios}.
Enormous set of results concerning structure and thermodynamics,
dynamics and dielectric properties was generated
~\cite{rao1,vaisman,luzar3,gunsteren1,bako1,jedlovszky1,samios,gunsteren2,skaf1,skaf2,vishnyakov,mancera,bagchi1,jedlovszky2,bagchi2,perera,jedlovszky3}.

At present, the most widely used models for the DMSO are at the united atom
level (DMSO-UA); in this framework, each methyl group is considered as a single
site. Nevertheless, the
attempts to improve the description of the properties of pure DMSO and water-DMSO
mixtures,  by using  either flexible or all-atom or 
simple polarizable DMSO models were
undertaken~\cite{bako1,strader,zuo,kollman,benjamin,senapati,zhao,bachmann}.
The majority of the MD studies for the united atom models
were performed at NVT conditions using experimental densities for different compositions
~\cite{rao1,vaisman,luzar3,gunsteren2,mancera}
or by means of NVE simulations~\cite{bako1,samios,skaf1,skaf2}.
The isobaric-isothermal (NPT) setup was employed 
as well~\cite{vishnyakov,bagchi1,bagchi2,perera,zuo,bordat,gujt}.
In a comprehensive study of excess mixing properties of water-DMSO model mixtures
at room temperature and atmospheric pressure, the NVT Monte Carlo technique
was applied~\cite{jedlovszky1}.
A peculiar behavior reflecting the hydrophobic effect in such mixtures 
was explored in~\cite{bagchi1,bagchi2}.
To summarize, the overwhelming body of computer simulation results refers to composition 
changes of the properties for the system in question rather than to temperature and/or pressure effect. 
Finally, it is worth mentioning that
the solvation of complex molecules in water-DMSO mixtures was the subject of 
several simulations and experimental studies, see e.g., 
references~\cite{bagchi3,martins,laria1,laria2,sokolowsky,kovalenko}.

\section{Models and simulation details of the present work}

In this work we restrict our attention to a few DMSO united atom type, non-polarizable  models 
with four sites, O, S, CH$_3$ (methyl group carbon
and hydrogens interaction sites are replaced by a single CH$_3$ site 
located on the methyl carbon atom). Within this type of modelling,
the interaction  potential between all atoms and/or groups is assumed 
as a sum of Lennard-Jones (LJ) and Coulomb terms. 
A comprehensive description of the parameters of models of this type
is given in table~1 of~\cite{mancera3} and in table~1  of~\cite{jedlovszky1}. 
In all cases, the models are considered as rigid.
Some basic properties under specific conditions
concerned with computer simulations setup are also reported 
in table~2 of~\cite{mancera3}. 
A set of parameters for each model follows from fitting to a few
target properties, e.g., the heat of vaporization and charge distribution in order to yield a
dipole moment of the molecule in the gas phase, which is the 
case to construct for example the P2 model for DMSO~\cite{luzar3}. Experimental predictions 
for the intra-molecular structure and auxiliary ab initio calculations were used in the design of DMSO models as well.
Several attempts to better describe a wider set of target properties of DMSO at
ambient conditions were undertaken via parametrization
of charges and LJ parameters still keeping the model rigid~\cite{gunsteren1,gunsteren2,bordat}.
In order to improve the rigid P2 model, Benjamin proposed to include the
effects of intra-molecular flexibility~\cite{benjamin}. The 
parameters for intra-molecular potential describing harmonic bonds and angles
were chosen to reproduce the gas phase normal mode frequencies. 
It was shown however~\cite{senapati}, that the model essentially overestimates 
the liquid DMSO density at room temperature. Thus, it is improbable to expect a
good level description of the density of water-DMSO mixtures over the 
entire composition range.

Both rigid or flexible DMSO model can also be constructed by using a well established
OPLS database~\cite{jorgensen}. The OPLS and P1 and P2 models~\cite{luzar3} 
are characterized by identical molecular geometry, though the LJ parameters are
different. The site charges are the same for the OPLS and P2 models.
In the construction of a flexible model, one can take parameters from the OPLS database.
However, to keep the molecular geometry, it is then recommended to complement
the DMSO model by  the improper dihedral angle S--C--O--C 
see reference~\cite{oostenbrink}.

In the present study we apply the  DMSO united atom models with four 
sites, O, S, CH$_3$. The CH$_3$ is abbreviated as C in what follows. The models are
non-polarizable and  denominated as P1, P2 ~\cite{luzar3} and OPLS (see e.g.,
table II of reference~\cite{skaf3}). The interaction potential between all atoms and/or groups 
is a sum of Lennard-Jones (LJ)  and Coulomb terms. All the parameters are given in table 1
for the sake of convenience of the reader.  In addition, after testing several possibilities, we
assumed C--S and S--O bonds rigid whereas the O--S--C and C--S--C are chosen flexible,
the harmonic constant is taken from the OPLS database. Moreover, the improper dihedral
S--C--O--C is included into the DMSO models. The parameters are taken 
from reference~\cite{oostenbrink}. Our tests indicate that this element is crucial
to yield a reasonable value for pure DMSO liquid density in each of the models 
involved in this study. In summary, we applied semi-flexible models for DMSO molecule.

  \begin{table}[h!]
  \small
    \caption{Lennard-Jones parameters for the P1, P2 and OPLS
    models of dimethylsulfoxide.
    All $\sigma$ are in \AA, $\varepsilon$ --- in kJ/mol and charges, $q$, in $e$ units.
    All models assume the O--S--C angle at 106.75$^{\circ}$
    and the C--S--C angle at 97.4$^{\circ}$. The charges are: 
    $q_{\rm{O}} = -0.459$, $q_{\rm{S}} = 0.139$ and $q_{\rm{C}} = 0.160$ for P2 and OPLS models.
    In the case of P1 model, the charges are:
    $q_{\rm{O}} = -0.54$, $q_{\rm{S}} = 0.54$ and $q_{\rm{C}} = 0$.}
    \label{T_1}
    \vspace{0.1cm}
     \begin{center}
      \begin{tabular}{l c c c c c c c c c }
      \hline
       Model  &    $\varepsilon_{\rm{OO}}$ & $\sigma_{\rm{OO}}$ &    $\varepsilon_{\rm{SS}}$ & $\sigma_{\rm{SS}}$ &
                 $\varepsilon_{\rm{CC}}$ & $\sigma_{\rm{CC}}$  \\
       \hline
       P1     &  0.2992 & 2.80 & 0.9974 &  3.40  & 1.230 & 3.80   \\
       P2     &  0.2992 & 2.80 & 0.9974 &  3.40  & 1.230 & 3.80   \\
      OPLS    &  1.171  & 2.93 & 1.652  &  3.56  & 0.669 & 3.81   \\
       \hline
  \end{tabular}
  \end{center}
\end{table}

For water, the TIP4P-2005 model was applied~\cite{carlos}. In many cases we used
the TIP4P/$\varepsilon$ model~\cite{alejandre} as well. It was designed to improve the
inadequacy of the prediction of the static dielectric constant of water by TIP4P-2005.
Lorentz-Berthelot combination rules were used to determine the cross parameters for
the relevant potential well depths and diameters for the P1-water and P2-water models,
whereas the geometric combination rule was assumed for the OPLS-TIP4P-2005 model. 

Molecular dynamics computer simulations of DMSO-water mixtures were performed in the
isothermal-isobaric (NPT) ensemble at atmospheric  pressure 1~bar and at temperature 298.15~K.
We used GROMACS package~\cite{gromacs} version 5.1.2.
The simulation box in each run was cubic, the total number of molecules of both species is
fixed at 3000. Composition of the mixture is described by the molar fraction of DMSO
molecules, $X_{d}=N_d/(N_d+N_w)$.
As common, periodic boundary conditions were used.
Temperature and pressure control was provided by the V-rescale thermostat and Parrinello-Rahman
barostat with $\tau_T = 0.5$~ps and $\tau_P = 2.0$~ps, the timestep was 0.001~ps.
The value of $4.5\times 10^{-5}$~bar$^{-1}$ was used for the compressibility of mixtures.

The non-bonded interactions were cut-off at 1.1~nm, whereas the long-range electrostatic interactions
were handled by the particle mesh Ewald method implemented in the GROMACS software package  (fourth
order, Fourier spacing equal to 0.12) with the precision $10^{-5}$.
The van der Waals correction terms to the energy and pressure were applied.
In order to maintain the geometry of water molecules and DMSO intra-molecular bonds rigid, the LINCS
algorithm was used.

After preprocessing and equilibration, consecutive simulation runs, each for not less than 10 ns,  with
the starting configuration being the last configuration from the previous
run, were performed to obtain trajectories for the data analysis. 
The results for each property  were obtained by averaging over 7--10 production runs.

\section{Results and discussion}

\subsection{Density of water-DMSO mixtures on composition}

As we mentioned in the introductory section, there were several
experimental reports concerning the evolution of density of DMSO-water mixtures 
upon changing composition. 
We used experimental data by Egorov and Makarov at room temperature
$T= 298.15$~K, and atmospheric pressure~\cite{egorov}.

\begin{figure}[h]
\begin{center}
\includegraphics[width=6.5cm,clip]{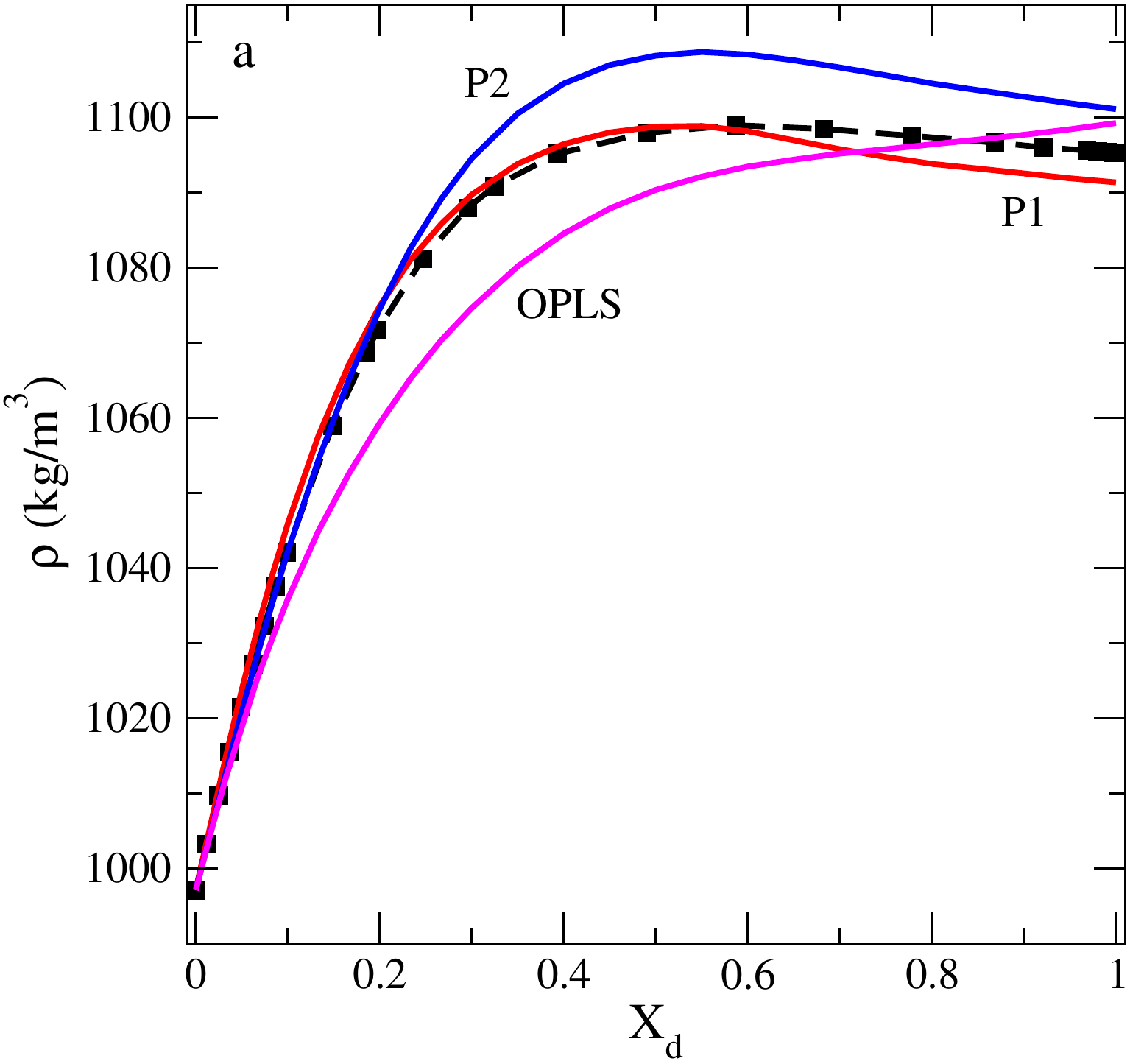}
\includegraphics[width=6.5cm,clip]{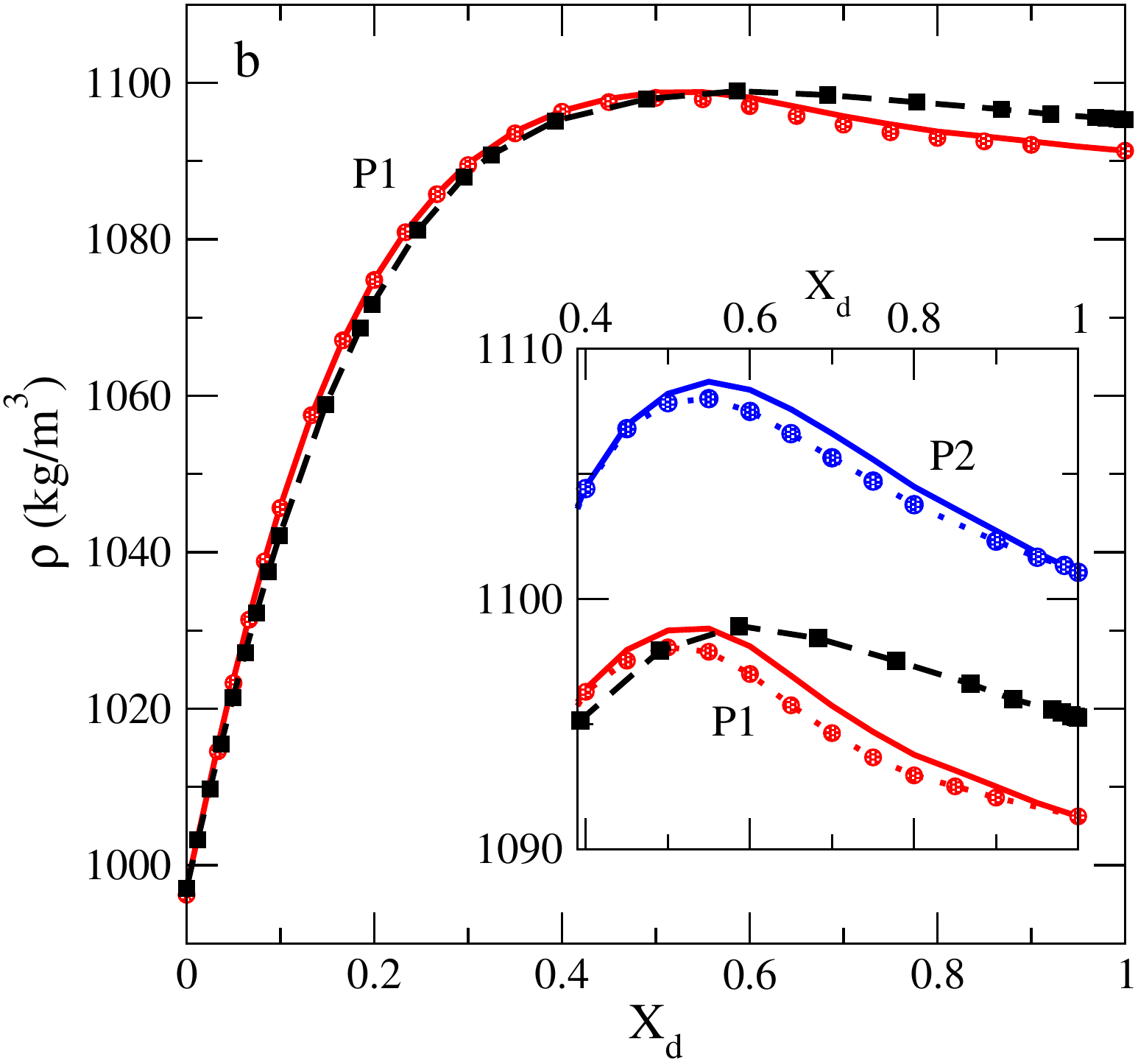}
\end{center}
\caption{(Colour online) Panel a: composition dependence of the density of water-DMSO 
mixtures from the NPT MD simulations of  P1-TIP4P-2005 model (solid red line),
P2-TIP4P-2005 model (solid blue line) and OPLS-TIP4P-2005 model (solid magenta line)
in comparison  with the experimental data (dashed lines with squares) at 298.15~K~\cite{egorov}.
Panel b: similar to the panel a but for P1-TIP4P-2005 model (solid red line), 
P1-TIP4P/$\varepsilon$  water model (solid circles). The inset gives an  enhanced view
of the effects of substitution of TIP4P-2005 (lines) by TIP4P/$\varepsilon$ (circles) water model
if the P1 and P2 models for DMSO are used. Experimental data are from ~\cite{egorov}.
}
\label{fig_1}
\protect
\end{figure}

The experimental data at $T=298.15$~K witness that the density of water-DMSO mixtures
increases quite rapidly with $X_{d}$ starting from pure water density ($997$~kg/m$^3$) and 
reaches a weakly pronounced maximum around $X_d \approx 0.6$. Next,  with a further
increase of $X_{d}$, i.e., in the interval where the DMSO amount within the mixture predominates, 
the density slightly decreases with $X_{d}$ and reaches the pure liquid DMSO 
density, $\rho = 1095 $~kg/m$^3$, figure~\ref{fig_1} (panel a). 
It is reasonable to interpret the presence of a maximum density along the composition axis
as an evidence of the stronger water-DMSO interaction, 
compared to the water-water inter-molecular interaction ~\cite{egorov}.

Molecular dynamics computer simulation results for P1, P2 and OPLS DMSO models
combined with TIP4P-2005 water model in the entire composition range are given 
in figure~\ref{fig_1}~a by red, blue and magenta lines, respectively. 
We observe that qualitative trends of behavior of density on
composition are correctly reproduced by the P1-TIP4P-2005 and P2-TIP4P-2005 models.
The maximum of density from simulations of these models is captured.
Moreover, the growth and decay of density in the interval of compositions with small DMSO amount 
and in the interval of small water concentrations, respectively, are reproduced
by both models as well.  By contrast, the OPLS-TIP4P-2005 model predicts a monotonous growth
of the mixture density with an increasing $X_d$ in the entire composition interval.
Thus, the simulations of all model combinations  show that the density increases upon adding
foreign species to water. If water is added to pure DMSO solvent, P1 and P2 predict a
growth of density. The OPLS model leads to an opposite trend, it yields a decreasing 
density, or in other words at constant pressure the volume of principal simulation box 
increases.  
If water enters DMSO structure within P2 or P1
models, the attraction dominates, then the density increases. 
On the other hand, if water enters the ``sea'' of OPLS molecules, 
repulsion prevails  and consequently the box increases or equivalently the density decreases. 

The P2-TIP4P-2005 model works very well up to $X_d \approx 0.2$. For this interval
of composition, the performance of the P1-TIP4P-2005 model is very satisfactory as well.
On the other hand, at higher values of $X_d$, the P2-TIP4P-2005 model overestimates 
the density of mixtures as the result of overestimation of pure DMSO liquid density. 
By contrast,  the P1 model slightly underestimates the density
of pure liquid DMSO. Consequently, the density of water-DMSO mixtures is slightly
underestimated at high $X_d$. 
In fact, this deviation in both cases 
is not too big. It is of the order of one percent w.r.t. experimental value.
If the TIP4P-2005 water model is substituted by the TIP4P/$\varepsilon$ model, the 
situation remains almost the same. In panel b of figure~\ref{fig_1}, we compare performance 
of the P2 model for DMSO  combined with two water models w.r.t. experimental data.
The inset to panel b of figure~\ref{fig_1} shows a marginal deviation of the mixture density 
upon changing the water model.  Furthermore, the inset shows that the maximum density from simulations
[$X_{d} \approx 0.55$ is close to its experimental counterpart ($X_{D} \approx 0.58$)].
In summary, the dependence of density on composition with a single
peculiar point is quite accurately described 
by two, P1 and P2, DMSO models if combined with either TIP4P-2005 or TIP4P/$\varepsilon$ models, 
whereas the OPLS-TIP4P-2005 model does not capture this behavior.

\subsection{Excess mixing volume, partial molar and apparent molar volumes of species on composition}

However, it is of importance not only to describe the trends of behavior of the given property 
on composition, but to correctly capture the deviation from ideality as well.
These insights follow from the behavior  of the excess 
density or the excess mixing volume, for example.
The excess mixing volume is defined as follows, $\Delta V_{\rm{mix}} = V_{\rm{mix}} - X_d V_d - (1-X_d) V_w$,
where $V_{\rm{mix}}$, $V_d$ and $V_w$ refer to the molar volume of the mixture and
of the individual components, DMSO and water, respectively.

\begin{figure}[h]
\begin{center}
\includegraphics[width=7cm,clip]{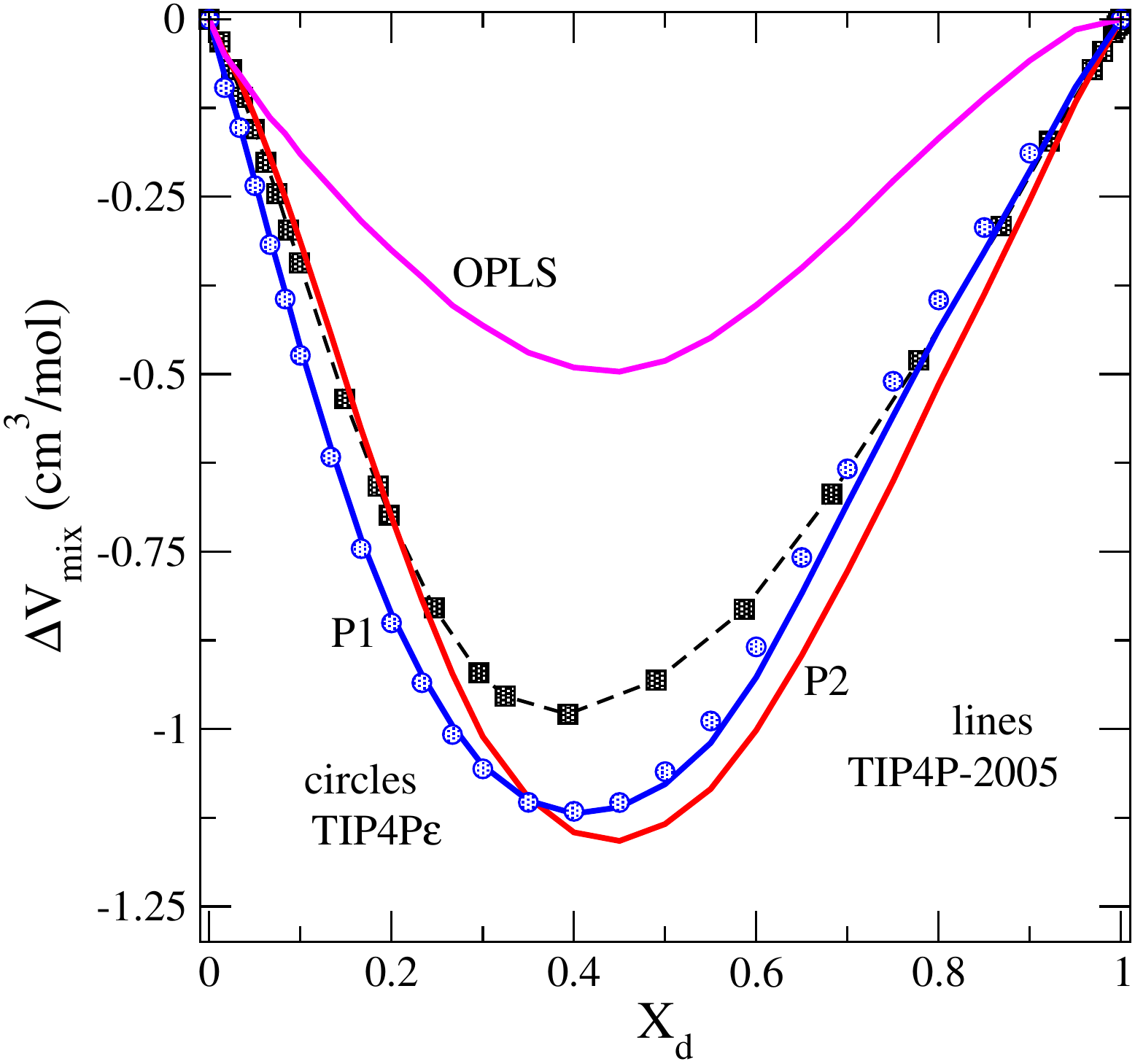}
\end{center}
\caption{(Colour online) A comparison of the composition dependence of the excess mixing volume
of water-DMSO mixtures for different DMSO (P1, P2 and OPLS) models with
the experimental data of reference~\cite{egorov}.
In addition, the effect of substitution
of TIP4P-2005 (blue line) by TIP4P/$\varepsilon$ (blue circles) water model is shown.
The results refer to  298.15~K and atmospheric pressure.
}
\label{fig_2}
\protect
\end{figure}

In order to elucidate the composition trends of the behavior of the excess
mixing volume, we plot the simulation results and experimental data at 298.15~K in figure~\ref{fig_2}. 
Experimental data show that $\Delta V_{\rm{mix}}$ is negative and exhibit
a minimum at $X_d \approx 0.4$. 
The simulation results qualitatively show similar trends of behavior. 
The values for $\Delta V_{\rm{mix}}$ are just slightly overestimated at intermediate compositions,
if P1-TIP4P-2005 or P2-TIP4P-2005 models are used. 
At low $X_d$ values, the P2-TIP4P-2005 model performs better than P1-TIP4P-2005 model. 
However, the P1-TIP4P-2005 model predicts the location of $\Delta V_{\rm{mix}}$ a bit better
than the  P2-TIP4P-2005. Also for high $X_d$ values,  P1-TIP4P-2005 is better 
than P2-TIP4P-2005. The OPLS-TIP4P-2005 essentially underestimates the excess
mixing volume in the entire composition interval.
Thus, a comparison between the experiment and simulations of P1 and P2
models with either TIP4P-2005 or TIP4P/$\varepsilon$ can be termed as quite satisfactory
concerning the geometric aspects of mixing of species.
In this projection of the equation of state for DMSO-water mixture, 
we observe a single peculiar point on composition 
in close similarity to the behavior of density, as expected.

\begin{figure}[h]
\begin{center}
\includegraphics[width=6.5cm,clip]{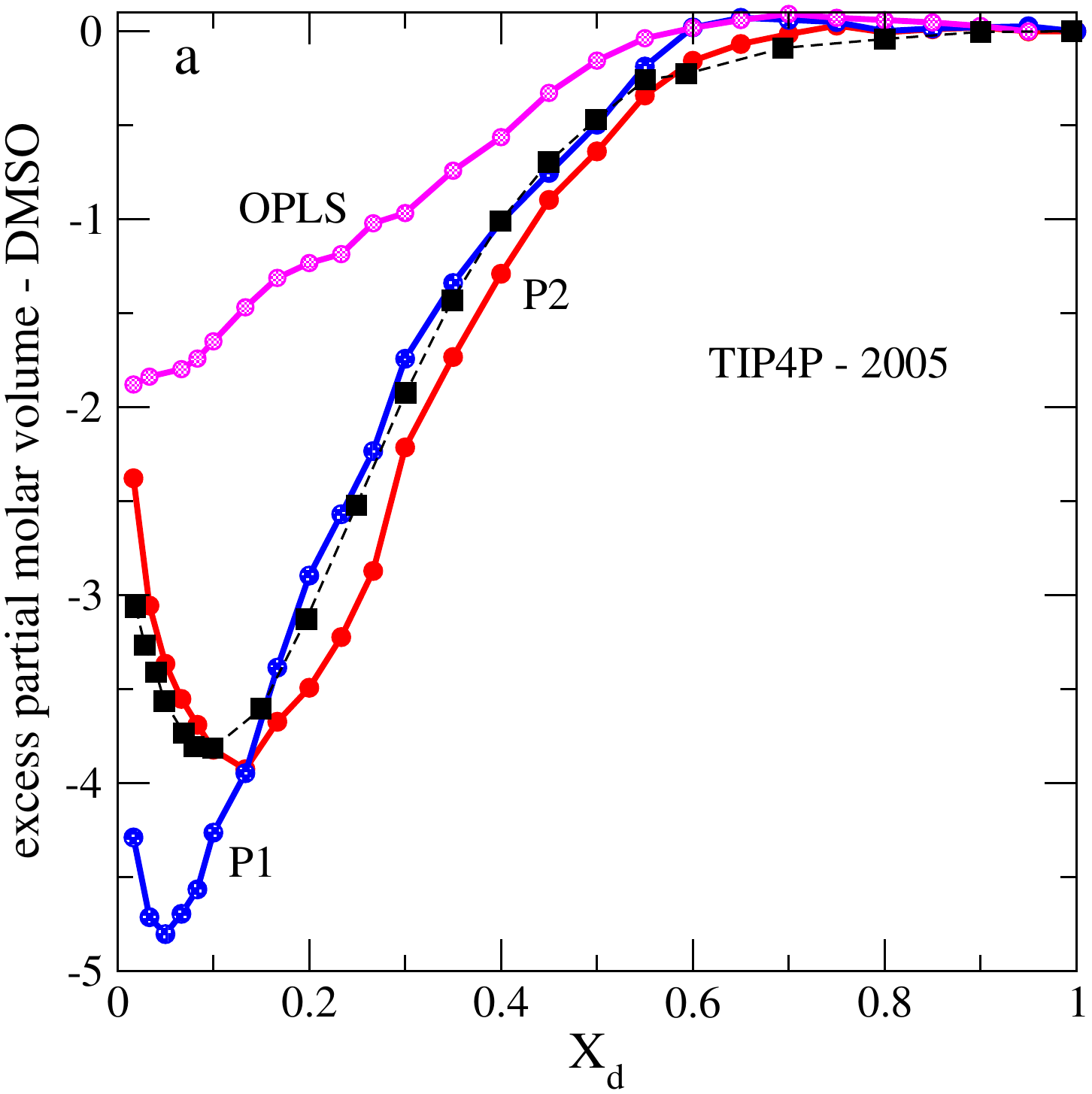}
\includegraphics[width=6.5cm,clip]{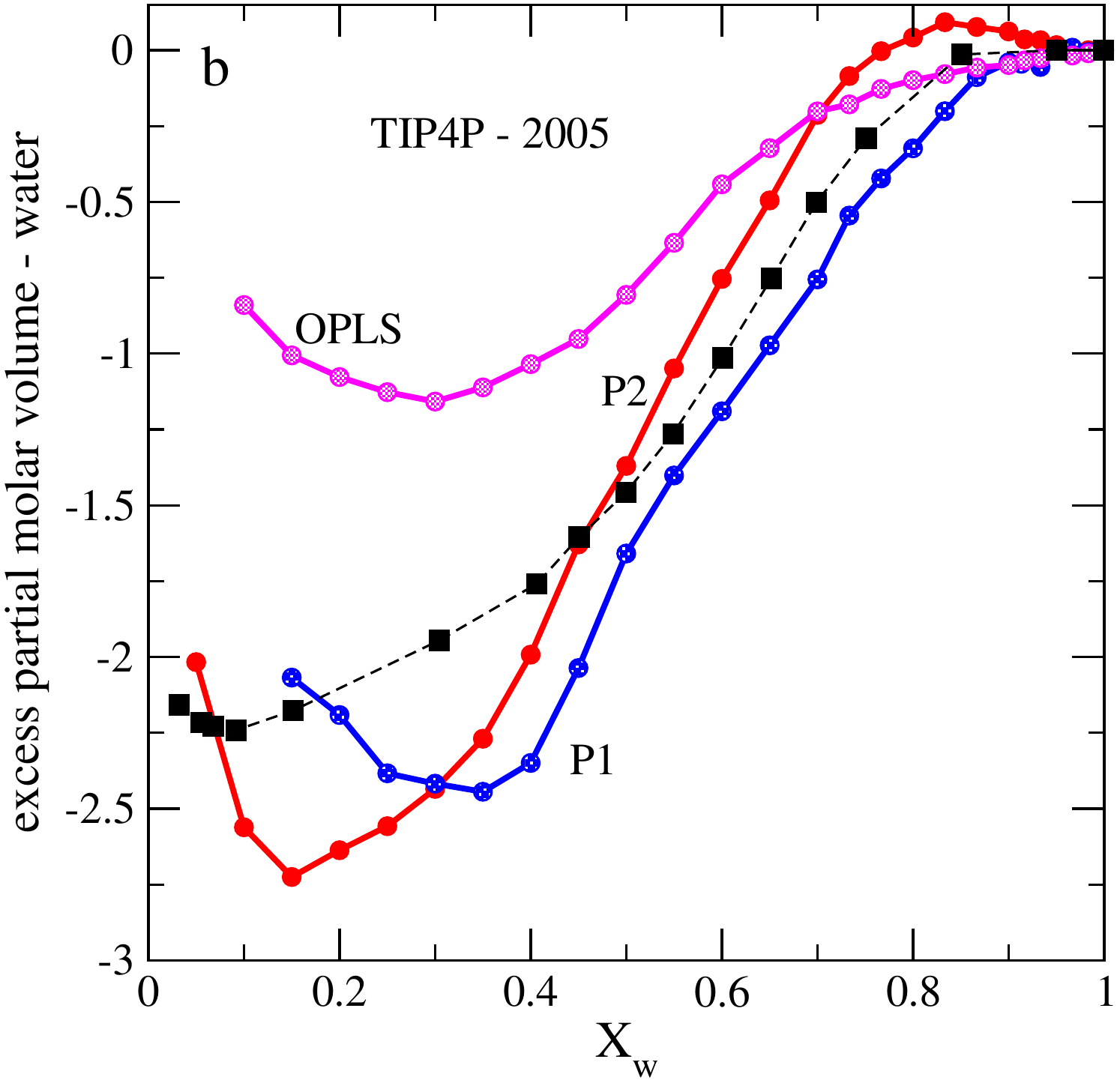}
\includegraphics[width=6.5cm,clip]{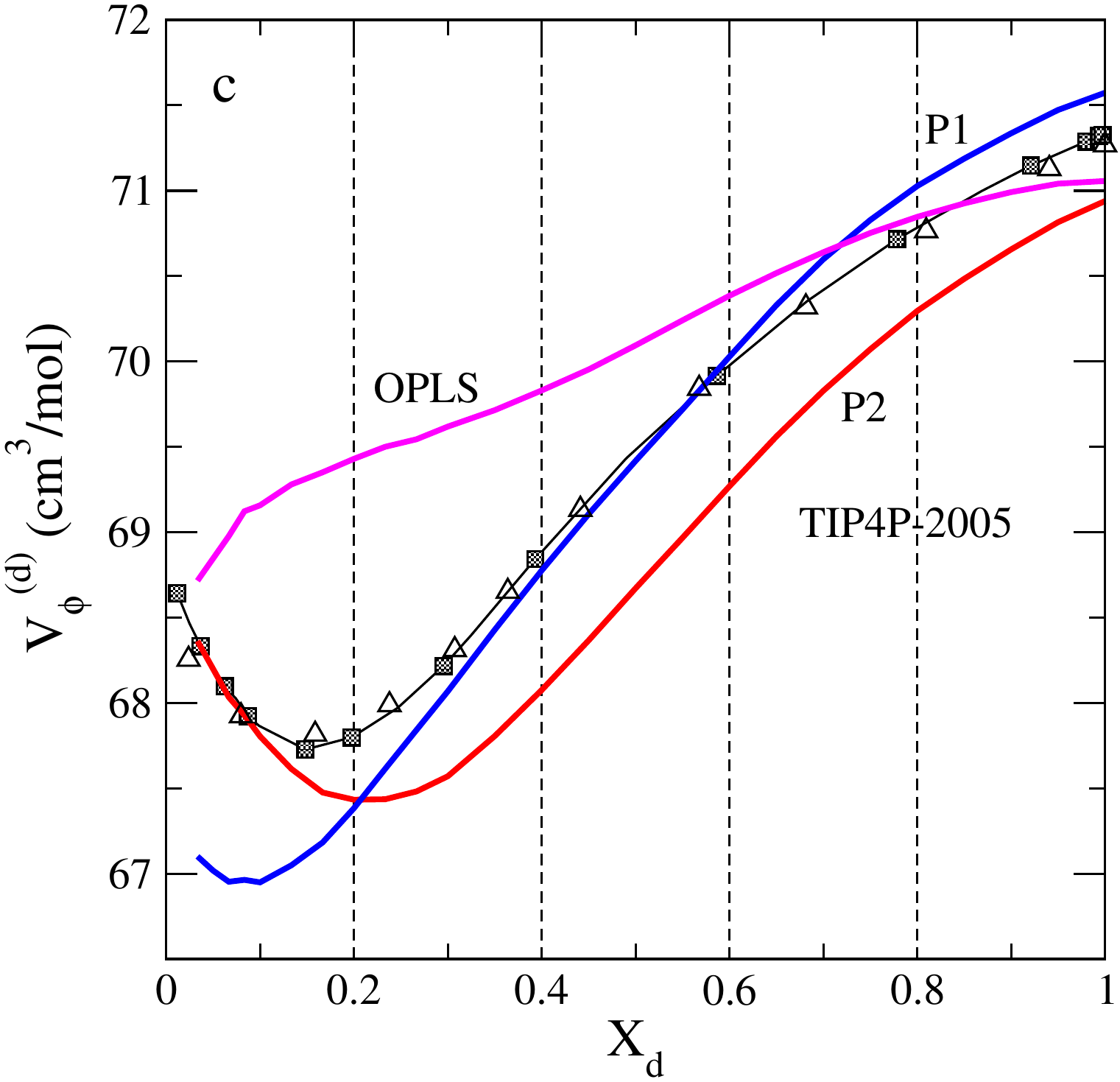}
\includegraphics[width=6.5cm,clip]{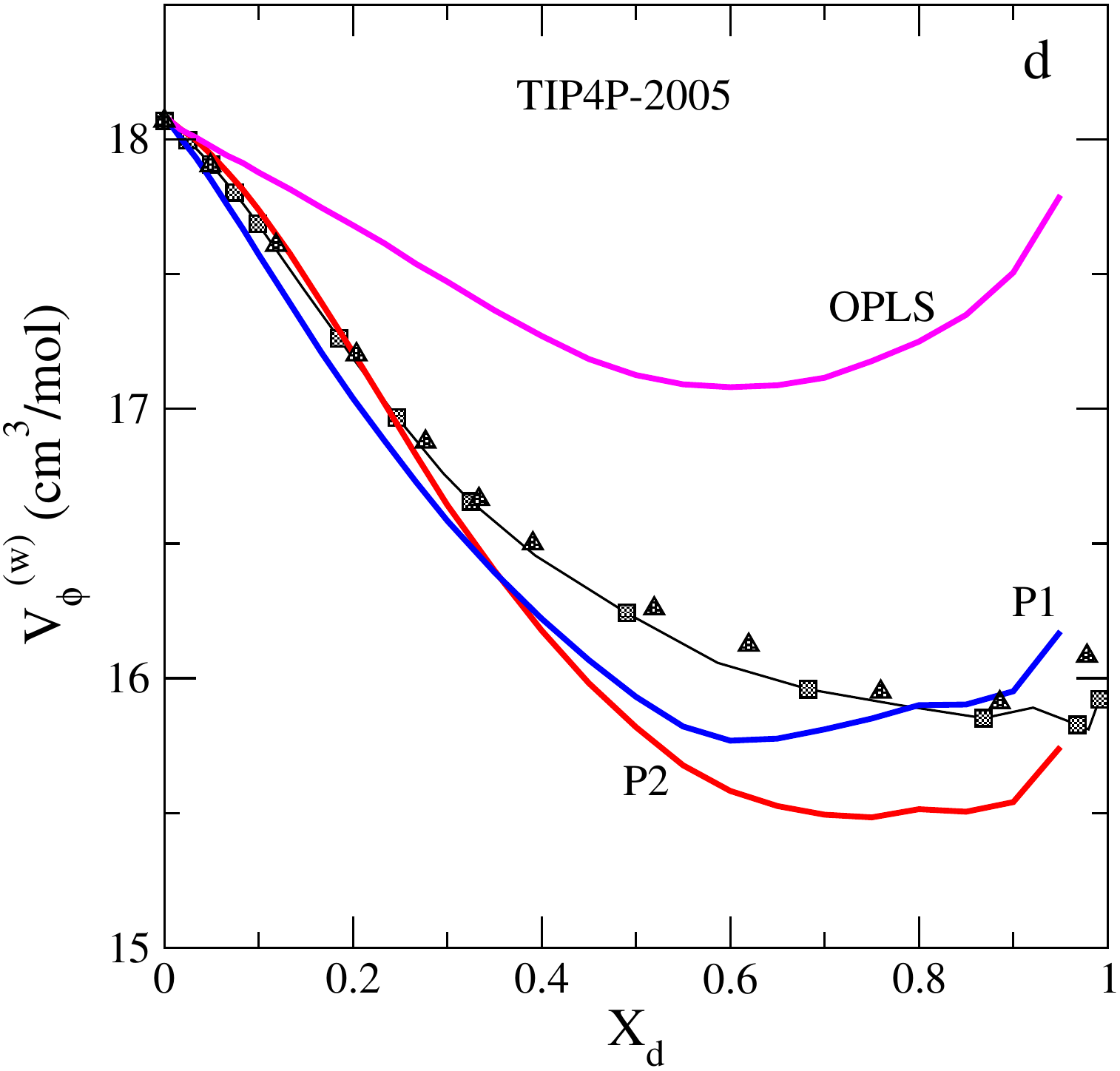}
\end{center}
\caption{(Colour online) Panels a and b: a comparison of the composition dependence of the excess
partial molar volumes of DMSO and of water from simulations,
with the experimental data (squares reference~\cite{torres}) at 298.15~K.
Panels c and d: a comparison of the composition dependence of the
apparent molar volumes of DMSO and of water from simulations,
with the experimental data (squares reference~\cite{egorov} and
triangles reference~\cite{carmen}) at 298.15~K.
}
\label{fig_3}
\protect
\end{figure}

The maximum deviation from the ideal mixing at $X_{D} \approx 0.4$ is
interpreted as trends for the formation
of ``complexes''  or ``associates''  of the 2DMSO-3H$_2$O type~\cite{vishnyakov}.
On the other hand, it is documented that the maximum deviation from ideality
of density and of other properties on composition
can be characterized in terms of percolation phenomena~\cite{perni}.

In order to discern the contribution of each species of the mixture into
the mixing volume, it is common to resort to the excess partial molar volumes.
They are defined as, $\overline{V^{ex}_i} = \overline{V}_i - V_i$,
where $\overline{V}_i$ denotes the partial molar volume of species $i$ ($w$ or $d$).
The partial molar volumes result from the excess mixing volume, $\Delta V_{\rm{mix}}$ as
follows~\cite{torres},
\begin{equation}
\overline{V}_w =\Delta V_{\rm{mix}}+V_w +X_d \left(\frac{\partial \Delta V_{\rm{mix}}}{\partial X_w}\right)\bigg|_{P,T},
\label{eq_3_1}
\end{equation}
\begin{equation}
\overline{V}_d =\Delta V_{\rm{mix}}+V_d -X_w\left(\frac{\partial \Delta V_{\rm{mix}}}{\partial X_w}\right) \bigg|_{P,T},
\label{eq_3_2}
\end{equation}
where $X_w = 1- X_d$.
The excess partial molar volumes of DMSO and water species are plotted in the panels
a and b of figure~\ref{fig_3}. The experimental data are taken from figures~5 and 6 of reference~\cite{torres}.
The best agreement with the experimental results for $\overline{V^{ex}_d}$ 
is obtained for the P2-TIP4P-2005 model. Namely, the shape of the curve and the minimum for
$\overline{V^{ex}_d}$ are reproduced quite well. Predictions from the  P1-TIP4P-2005 model
are reasonable as well. However, the location and the value for the minimum are less
accurate in comparison to P2-TIP4P-2005 model. 
Concerning the predictions of different models w.r.t. the $\overline{V^{ex}_w}$, the situation
is a bit worse. Both models, the  P2-TIP4P-2005 and  P1-TIP4P-2005, describe the
behavior of the excess partial molar volume well at a high water content. On the other hand,
for DMSO-rich mixtures, the simulation data deviate from the experimental results.
The experimental data show that the minimum of $\overline{V^{ex}_w}$ is weakly pronounced 
and is located at $X_w \approx 0.1$. By contrast, the simulation data predict much stronger
pronounced minimum at a higher water content. In summary, the dissolution of water species in
the DMSO solvent is not accurately described. The worse predictions for $\overline{V^{ex}_d}$
and $\overline{V^{ex}_w}$ come out from the application of the OPLS-TIP4P-2005 model.

Similar insights into the geometric aspects of mixing on composition,
both from experiments and simulations, can be obtained by resorting to the notion of
the apparent molar volume of species rather than the partial molar volumes.
The apparent molar volume
for each species~\cite{torres} according to the definition is:
$V_{\phi}^{(w)}= V_w + \Delta V_{\rm{mix}}/(1-X_d)$ and  $V_{\phi}^{(d)}= V_d + \Delta V_{\rm{mix}}/X_d$.
We elaborated the experimental density data from references~\cite{egorov,carmen}
and the results from our simulations to construct the plots shown in panels c and d of figure~\ref{fig_3}.
These plots confirm that the P2-TIP4P-2005 model provides a reasonable description of the
composition behavior for $V_{\phi}^{(d)}$ in water-rich mixtures while the P1-TIP4P-2005
model is satisfactory for DMSO-rich mixtures.

This kind of composition behavior of apparent molar volumes of DMSO species 
in water-rich mixtures can be related to the experimental results for abnormal 
intensity of scattered light~\cite{rodnikova}. The minimum of the apparent
molar volume of DMSO indicates the hydrophobic effect in the systems in question
due to enhanced concentration fluctuations within certain composition interval.

\subsection{Energetic aspects of mixing of DMSO and water molecules}

Next, we proceed to the energetic manifestation of mixing trends. 
The experimental data are scarce in this aspect, we used the results from
references~\cite{clever,rodante,lai,cowie} to explore the mixing enthalpy upon composition of the
water-DMSO mixtures. 

\begin{figure}[h]
\begin{center}
\includegraphics[width=6.5cm,clip]{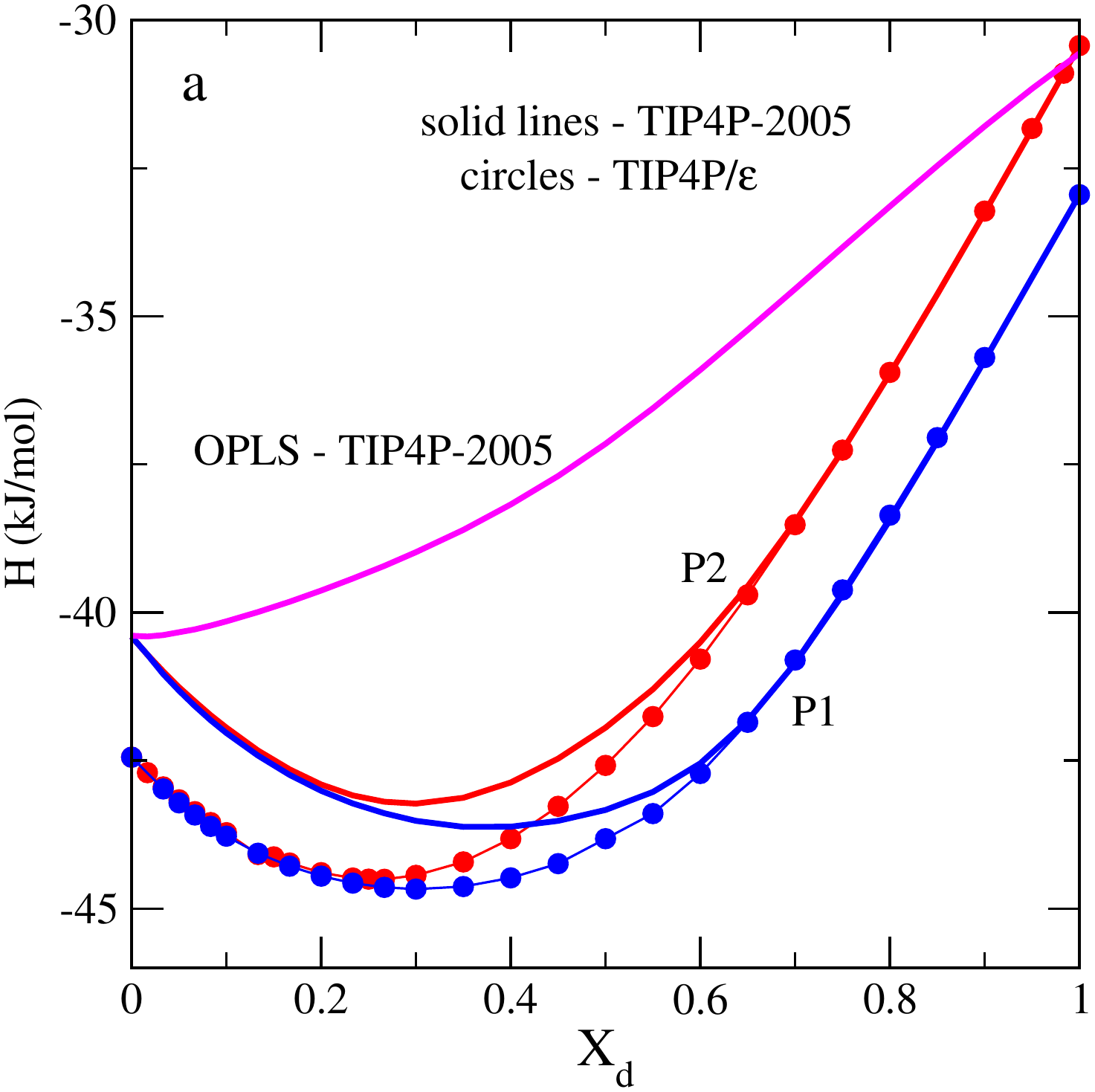}
\includegraphics[width=6.5cm,clip]{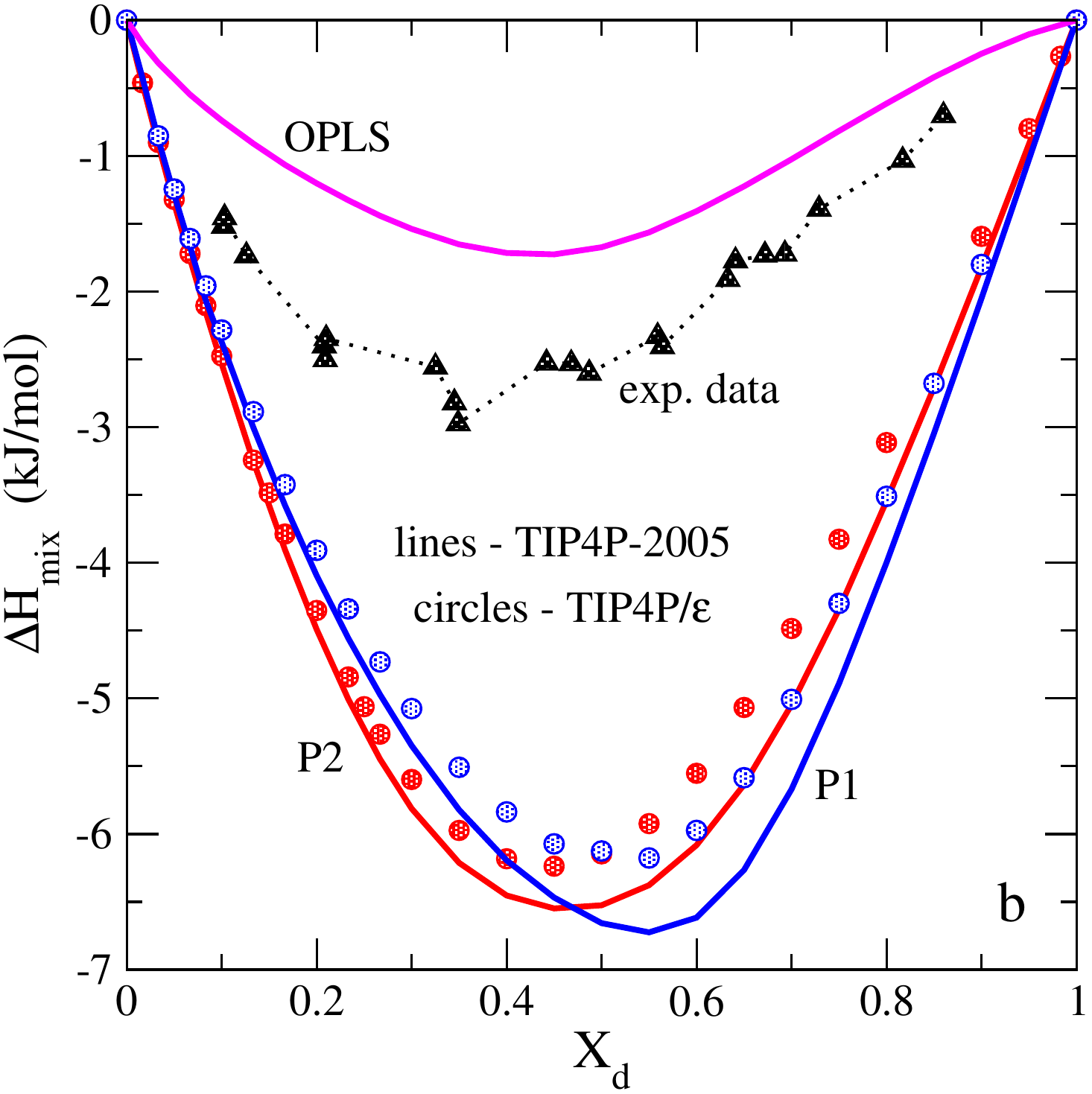}
\includegraphics[width=6.5cm,clip]{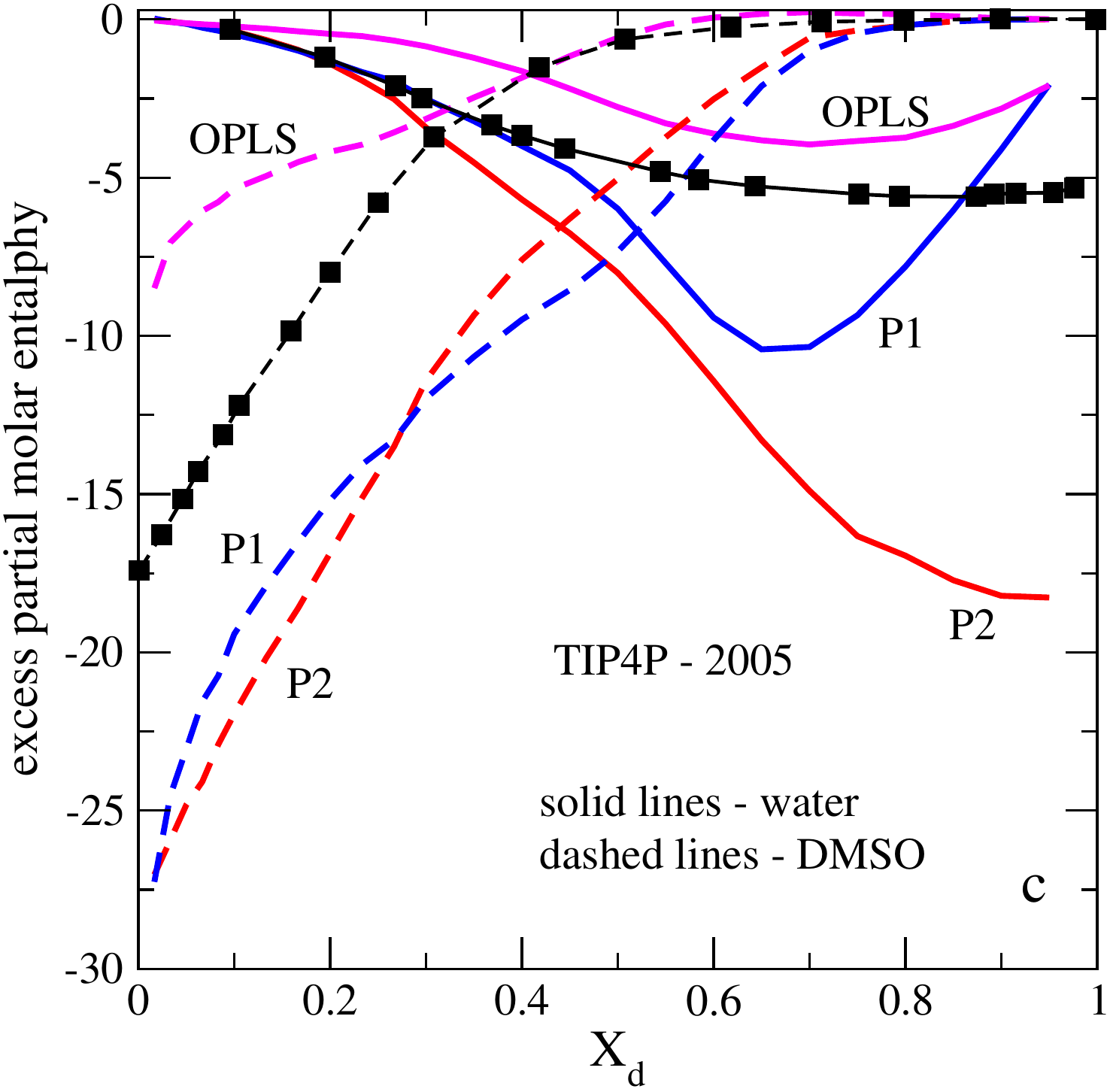}
\end{center}
\caption{(Colour online) Panel a: composition dependence of enthalpy of water-DMSO mixtures
at  298.15~K for different combinations of DMSO and water models.
Panel b: a comparison of the behavior of the excess mixing enthalpy
from simulations of different models and experimental data from refe\-rence~\cite{clever}.
The nomenclature of lines and symbols is given in the figure.
Panel c: excess partial molar enthalpies of species on composition from simulations.
The experimental data are from Lai et al.~\cite{lai}.
}
\label{fig_4}
\protect
\end{figure}

To begin with, the simulation results for the enthalpy on composition for different combinations
of models are given in panel a of figure~\ref{fig_4}. The curves resulting from  P2-TIP4P-2005 and
P1-TIP4P-2005 predict the presence of a minimum at $X_d \approx0.35$ and at 0.45, respectively.
By contrast, the OPLS-TIP4P-2005 does not predict the existence of the minimum for $H(X_d)$. If
one applies the TIP4P/$\varepsilon$ model rather than the TIP4P-2005 model, the absolute
values of enthalpy increase and the minimum shifts to a lower DMSO concentration. 

The 
experimental results, however, refer to the excess mixing enthalpy rather that to the absolute values.
It follows from the results shown in panel b of figure~\ref{fig_4}, that the P1 and P2 DMSO models
combined with the TIP4P-2005 water substantially over-predict the mixing enthalpy. 
Changing the water model from TIP4P-2005 to TIP4P/$\varepsilon$ does not yield an improvement
in the considered aspect. This seems to be a quite general
failure of a set of non-polarizable water and DMSO
combinations of models as documented in the comprehensive study~\cite{jedlovszky1},
see figures~3 and 5 of that reference.
On the other hand, from our figure~\ref{fig_4}~b, we observe that the OPLS-TIP4P-2005 model 
underestimates the mixing enthalpy but
the results are not far from the experimental points.  The excess partial molar enthalpies
can be obtained by using equations~(\ref{eq_3_1}) and (\ref{eq_3_2}), just the volume should be substituted by enthalpy.
The agreement between the experimental data and predictions of a set of models in
question can be termed as qualitative and in some cases as quantitative for water-rich 
or DMSO-rich mixtures. The OPLS-TIP4P-2005 model yields the most reasonable coincidence
with the experimental data. However, the energetic aspects of mixing of species,
in terms of $\overline{H^{ex}_w}$ and $\overline{H^{ex}_d}$ simultaneously,
are better described for water-rich mixtures.

\subsection{Comments on the evolution of the microscopic structure upon changes of composition}

In all the above discussion concerned with geometric and energetic aspect of mixing of
water and DMSO species from molecular dynamics computer simulations of 
different models, we have intended to validate the theoretical predictions
w.r.t. experimental results in every detail. One of well established observations 
is in the existence of the minimum of the excess partial molar volume
or of the apparent molar volume of DMSO species at a low $X_d$.  This specific
interval of composition is the subject of search of anomalous behavior of 
different properties in references~\cite{bagchi1,bagchi2,bagchi3}. Some  
insights in those works were obtained from the analysis of the microscopic structure. 

Without referring directly to experimental results, it is difficult to prove that the 
computer simulation predictions for the
microscopic structure of the mixtures in question are reasonable. 
Recently, the contributions of our collaborators~\cite{pusztai1,galicia} 
have scrutinized this problem for pure water
and water-methanol mixtures. Unfortunately, we are not aware of 
comprehensive experimental measurements of the total and partial structure factors 
in the entire composition interval for water-DMSO mixtures. 
Thus, the discussion below should be considered as mirror interpretation of thermodynamic results
given above, rather than as independent and confident source to capture possible 
anomalous behaviors.

Consequently, we would like to perform mere analyses of the structural motifs
by considering how the partial radial distribution functions change upon adding
the DMSO molecules to water. The results are presented in the following figures.
We restrict principally to the results for P2-TIP4P-2005 model because it provides
the best description of the density dependence on composition and of other properties
of water-rich mixtures, as documented above.

\begin{figure}[h]
\begin{center}
\includegraphics[width=6.5cm,clip]{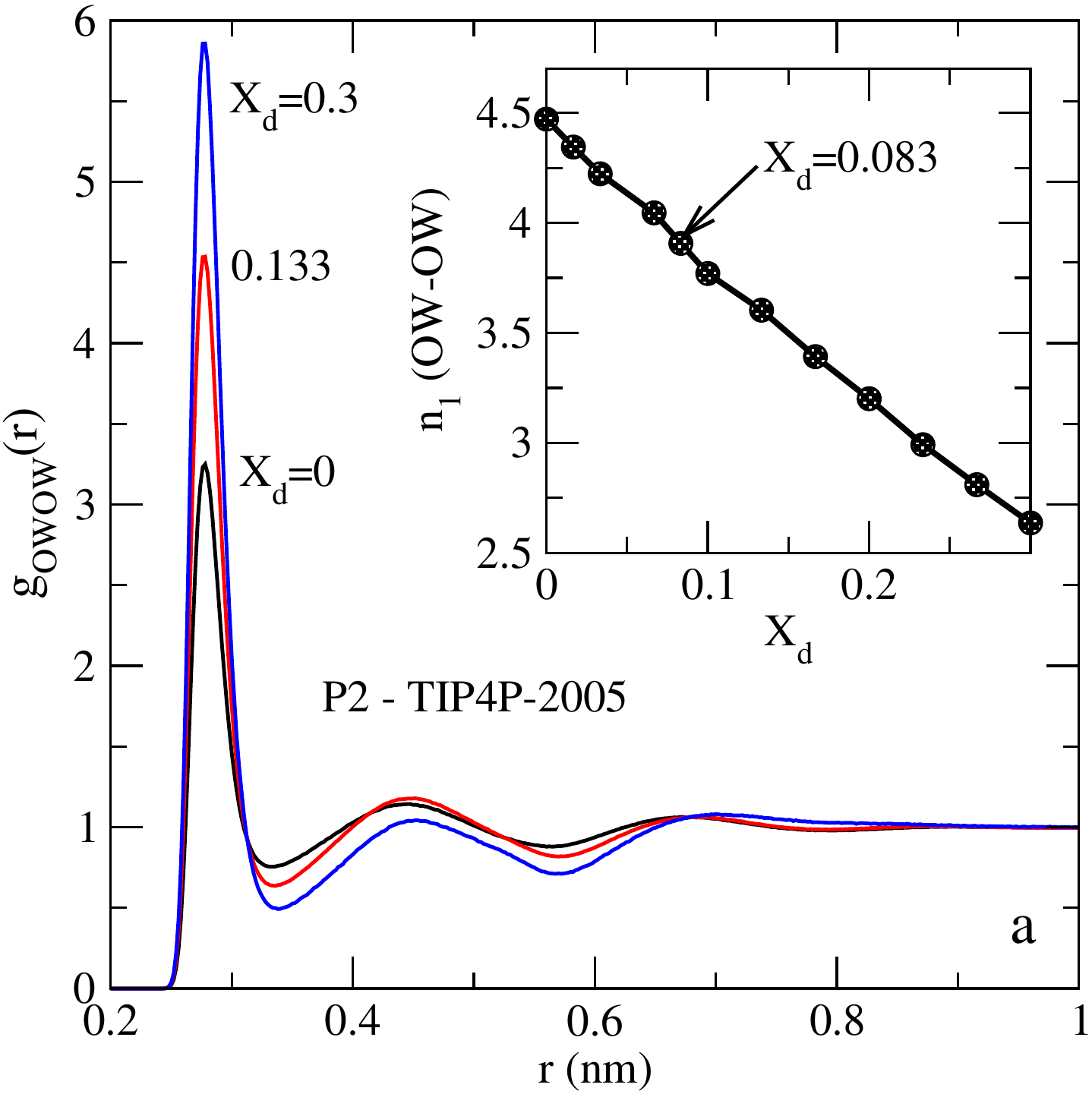}
\includegraphics[width=6.5cm,clip]{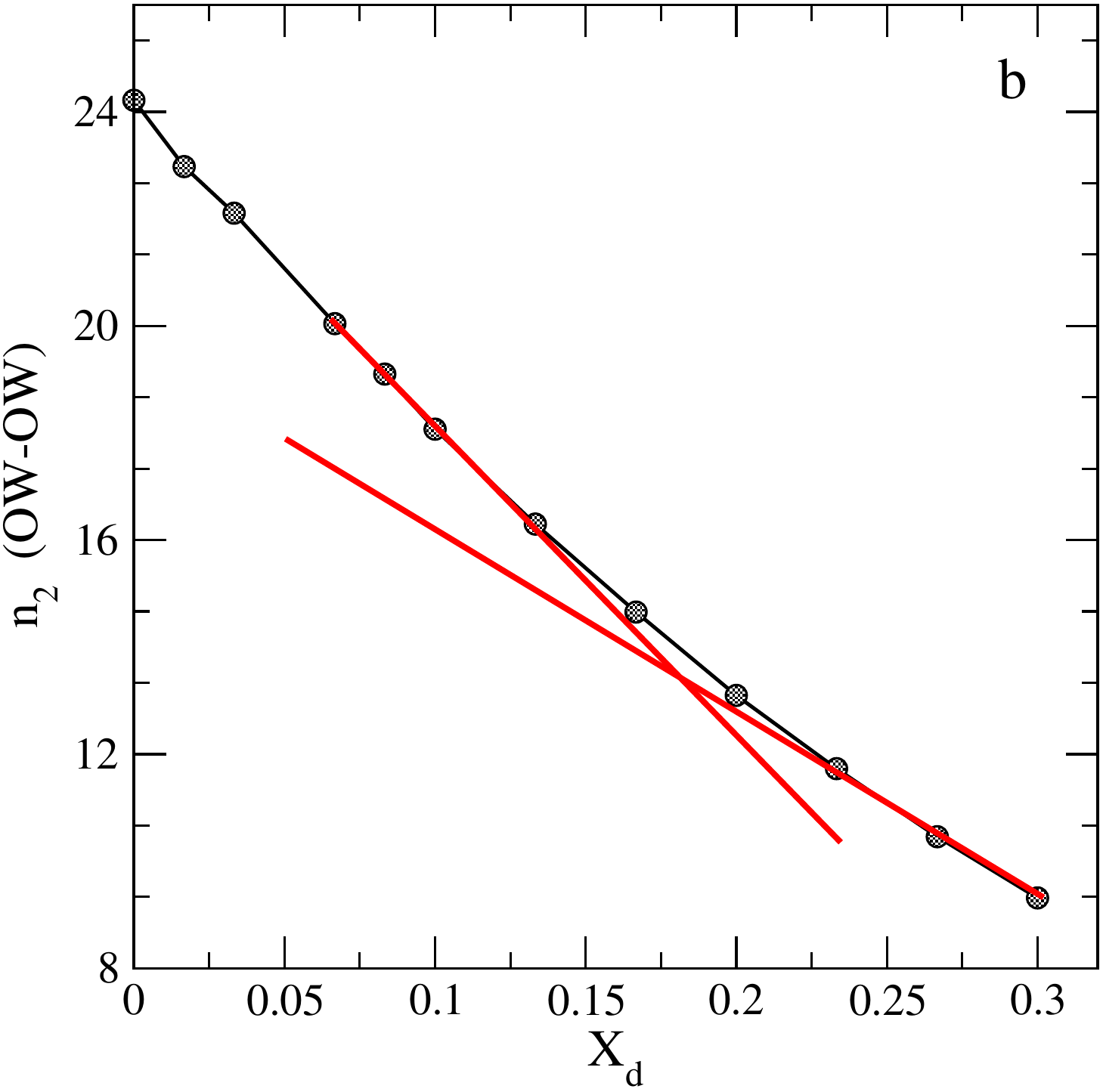}
\end{center}
\caption{(Colour online) Panel a: composition dependence of
the radial distribution functions of water oxygens for P2-TIP4P-2005 model
in water-rich mixtures. Inset: the first coordination number on DMSO concentration.
Panel b: evolution of the second coordination number upon mixture composition.
}
\label{fig_5}
\protect
\end{figure}

\begin{figure}[h]
\begin{center}
\includegraphics[width=6.5cm,clip]{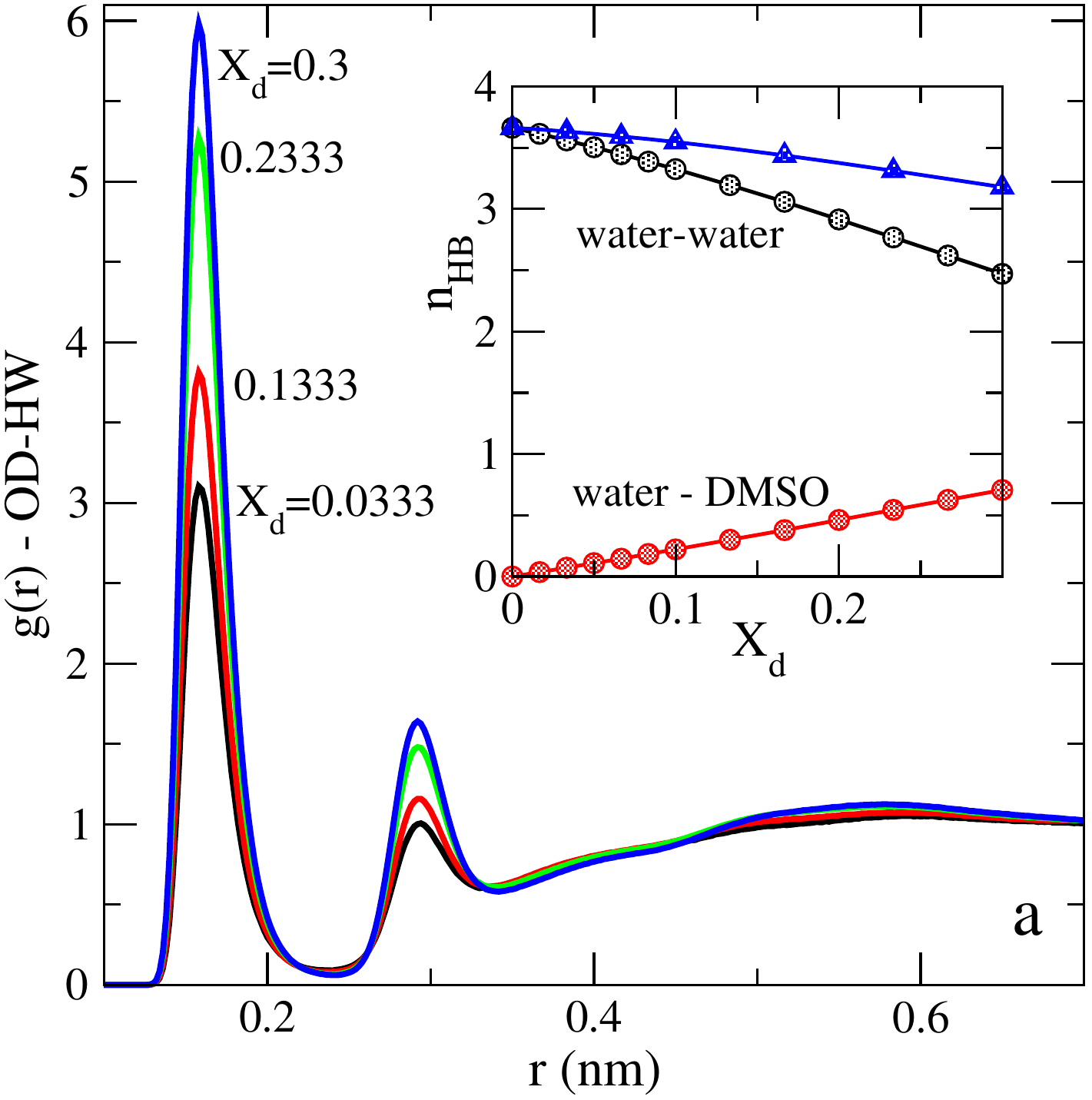}
\includegraphics[width=6.5cm,clip]{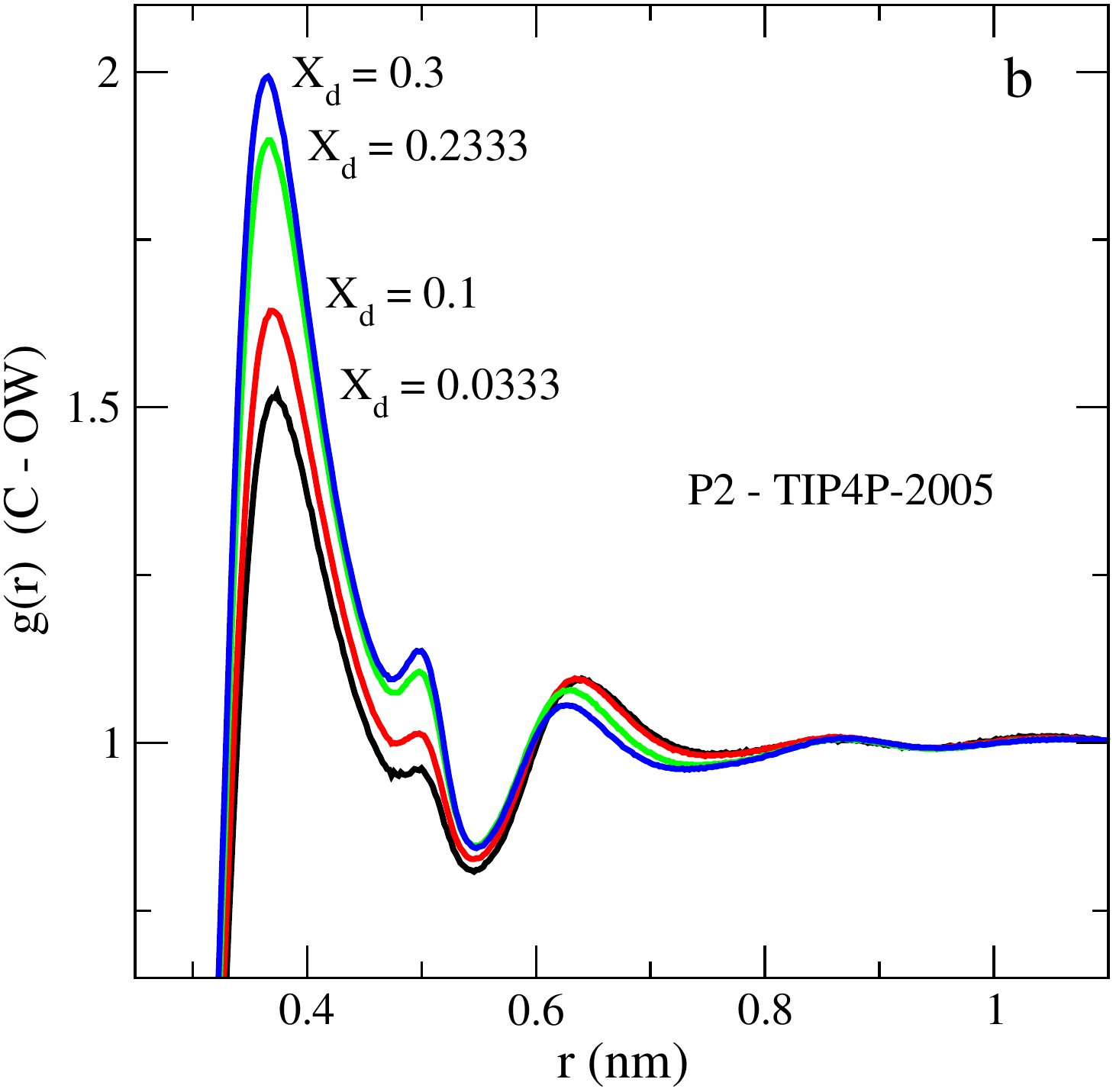}
\end{center}
\caption{(Colour online) Composition dependence of the cross radial distribution functions
between atoms of DMSO and water molecules. The inset to panel a shows
changes of the average number of hydrogen bonds per water molecule
between water molecules (black circles),
between water and DMSO molecules (red circles) and their sum (blue triangles).
}
\label{fig_6}
\protect
\end{figure}

\begin{figure}[h]
\begin{center}
\includegraphics[width=6.5cm,clip]{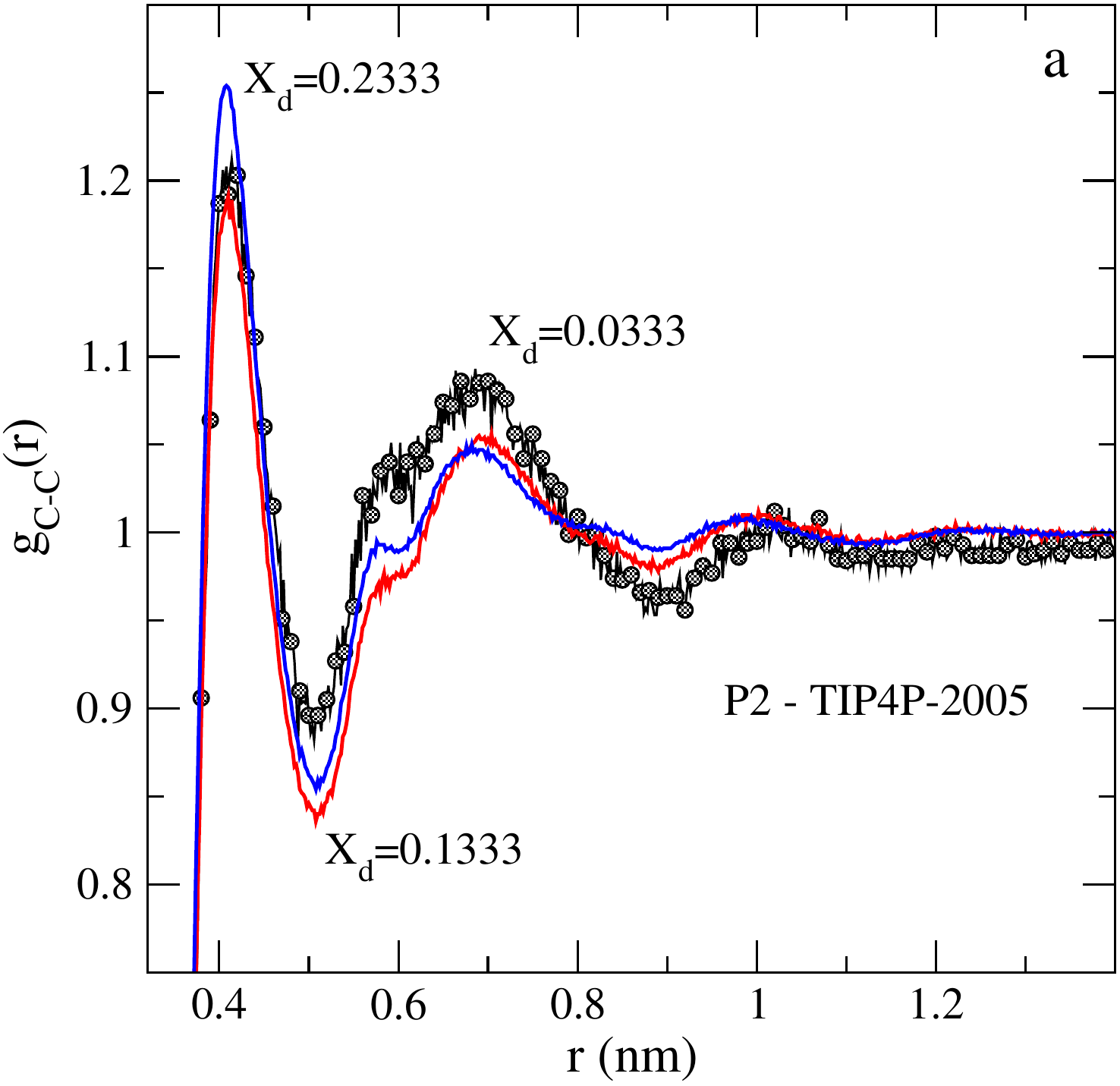}
\includegraphics[width=6.5cm,clip]{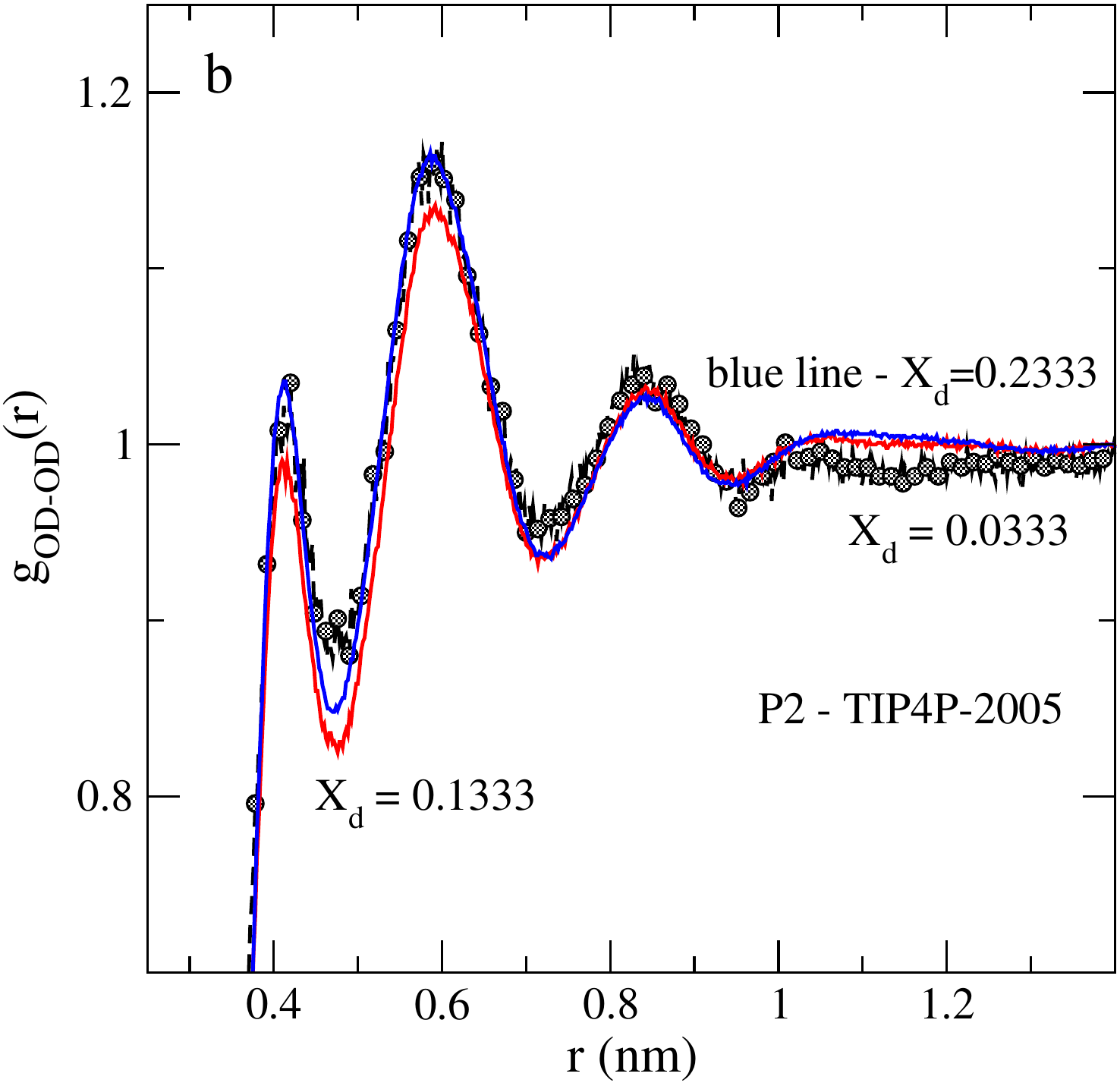}
\end{center}
\caption{(Colour online) Composition dependence of
the radial distribution functions of selected species for P2-TIP4P-2005 model
in water-rich mixtures. 
}
\label{fig_7}
\protect
\end{figure}

We observe that some of the functions exhibit anomalous changes whereas the other functions
change monotonously. Specifically, the radial distribution function for water oxygens
in given in figure~\ref{fig_5}. The first maximum of this function grows in magnitude with an increasing
DMSO concentration. This growth is accompanied by a decrease of the height of the second maximum.
Moreover, the location of the first maximum along $r$-axis does not change with $X_d$, whereas the
following oscillations slightly shift to larger distances. These changes can be characterized
by the first (inset to panel~a) and the second (panel~b) coordination numbers evaluated by integration
to the respective minimum of the $g(r)$ as common,
\begin{equation} 
  n_{i}(r) = 4\piup \rho_j \int_0 ^{r_{\rm{min}}}  g_{ij}(R)  R^2 \rd R,
\end{equation}
where $\rho_j$ is the density of species $j$.

This kind of evolution of $g(r)$ for water
species in water-DMSO mixtures was discussed already in reference~\cite{luzar3} (figure~7 of ~\cite{luzar3}) 
and in reference~\cite{gujt}. In addition to previous observations, we would like to note
that the first coordination number exhibits a weak nonlinearity as shown in the inset to figure~\ref{fig_5}.
Besides, we observed that the second coordination number changes its shape on $X_d$ at
$X_d \approx 0.1667$ as a reflection of changes of the second maximum of $g(r)$ for water
oxygens. A narrow range of compositions around this value is analyzed in~\cite{bagchi1}
suspecting an anomalous behavior of the number of contacts between methyl groups of DMSO molecules
in water-rich mixtures, cf. figures~8 and 9 of reference~\cite{bagchi1}.
In summary to figure~\ref{fig_5}, we note that the behavior of $g(r)$ for OW-OW can be interpreted 
as the enhancement of the short-range structure of water upon adding the DMSO species.
On the other hand, trends for a weaker structure at larger separations between water
molecules witness that on a larger scale, hydrogen bonded network of water becomes dismantled. 
Nevertheless, the entire interval of composition considered in figures~\ref{fig_5} and~\ref{fig_6} refers to 
systems with a rather high degree of hydrogen bonding. The average number of hydrogen
bonds (per water molecule) between species of the mixture is illustrated in the inset
to figure~\ref{fig_6}. We used GROMACS software with default parameters to obtain the data.
The average number of H-bonds between water molecule drops down from 3.7 to 2.5 within
the $X_d$ interval considered. On the other hand, the bonds are formed upon DMSO
insertion into water medium, such that their sum, or say bonding state of water molecules,
remains almost as high as in pure water. We have not elaborated the H-bonding structure
of the systems in question more in detail, however. Certain additional details can be 
deduced by using a previous observation for our laboratory, see e.g., figure~8 and 9 
of reference~\cite{gujt}. It is worth to extend the study of these issues by involving the
energetic definition of hydrogen bonding,
see e.g.,~\cite{bako2}, and by searching for the elements of microscopic structure
due to cooperativity of hydrogen bonds in the spirit of reference~\cite{pusztai2}.  
The hydrogen bonds lifetime should be explored as well.

Some aspects of the structural arrangement of different species in water-DMSO mixtures
upon increasing the DMSO concentrations are illustrated in figure~\ref{fig_6}. The principal effect
is shown in panel~a of this figure. Namely, the probability of finding water hydrogens
close to the oxygens of DMSO molecules substantially increases with an increasing $X_d$.
The second maximum increases as well. The first maximum is observed at the same OD-HW 
distance upon increasing $X_d$. Thus, this structure reflects the formation of hydrogen
bonds between DMSO and water molecules. The second maximum, presumably corresponds to the 
second shell of water molecules  bonded to molecules of the first shell around the ``reference''
oxygen of DMSO molecule. At larger distances, the distribution function remains almost flat.  
A quite similar shape of this function was documented previously
at $X_d=0.35$, cf. figure~3 of reference~\cite{luzar3}.

Because of bonding discussed just above, the distribution function of water oxygens w.r.t. 
methyl groups exhibits similar trends, panel~b of figure~\ref{fig_6}. Its first maximum increases with
increasing $X_d$. However, the second maximum  remains constant and then decreases in magnitude
when the DMSO concentration increases. Note that both maxima are observed at the distances larger
than the first and the second maximum of OD-HW distribution in panel~a, as expected.

The pair distribution functions describing the changes of distribution of methyl
groups, C-C (the inter-molecular part) and of OD-OD  behave differently, in comparison 
to our discussion above. They are plotted in figure~\ref{fig_7}. 
Namely, the first maximum of the radial distributions
shown in panels~a and~b decreases in magnitude when $X_d$ increases from $0.0333$ up to
$0.1333$. Next, the first maximum of both functions grows, if $X_d$ increases further, up 
to $0.2333$. The magnitude of the second maximum exhibits similar trends.
The shape of the functions from our calculations favourably compares to the ones
obtained in~\cite{luzar3} by using the P1 and P2 models
combined with the SPC water model, cf. figures~1 and 4 of that reference.

\begin{figure}[htb]
\begin{center}
\includegraphics[width=6.5cm,clip]{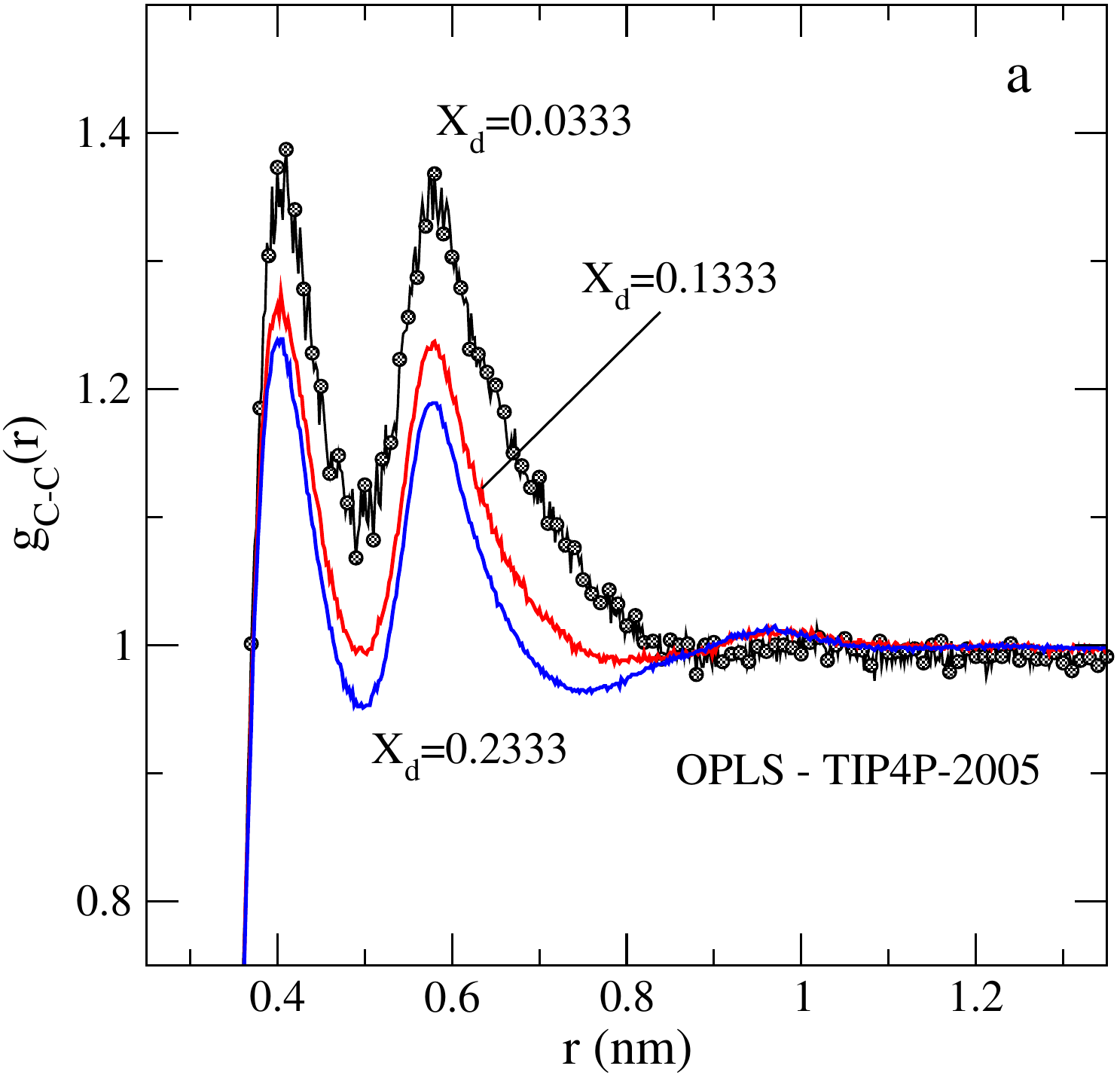}
\includegraphics[width=6.5cm,clip]{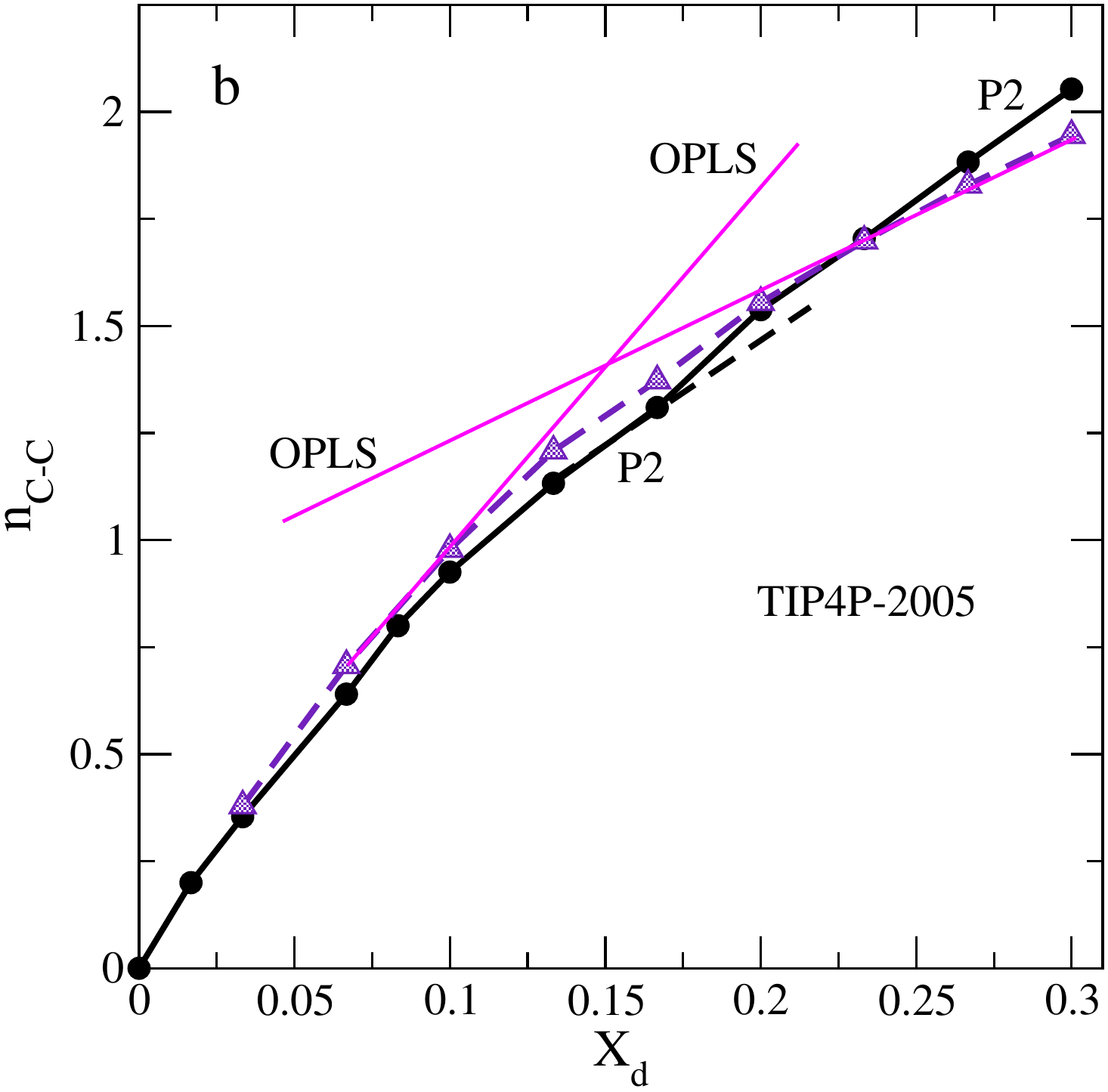}
\end{center}
\caption{(Colour online) Panel a: evolution of the pair distribution function of methyl groups
for the OPLS-TIP4P-2005 model at different compositions $X_d$.
Panel b: coordination number of a single methyl group on composition changes 
of water-DMSO mixtures from the OPLS-TIP4P-2005 and P2-TIP4P-2005 models.
}
\label{fig_8}
\protect
\end{figure}

However, the radial distribution functions of methyl groups require additional 
comments because it should reflect possible aggregation of hydrophobic methyl groups.
In order to get more detailed insights into the behavior of this function,
we elaborated figure~\ref{fig_8}~a by using the OPLS-TIP4P-2005 model for mixtures in question.
The shape of the curves differ from what we obtained for P2-TIP4P-2005 model in
figure~\ref{fig_7}~a. However, in qualitative terms, the C-C function from the OPLS-TIP4P-2005 model
is very similar to the result of Bagchi et al. in reference~\cite{bagchi1} using
another force field, cf. figure~6 of ~\cite{bagchi1}.
  
We used additional descriptors to
characterize the distribution function of methyl groups. Namely, we calculated
the first coordination number of the distribution of a single methyl group
of a DMSO molecule, figure~\ref{fig_8}~b,
using the P2-TIP4P-2005 and OPLS-TIP4P-2005 model. Apparently,
both curves are smooth without any discontinuity.
The inclination of the curve coming from OPLS-TIP4P-2005 model continuously changes 
in the composition interval between 0.1
and 0.2. The ``crossover''  between parts of the function on the DMSO concentration 
can be estimated at $X_d$ = 0.15.
A similar value for a peculiar composition was obtained in reference~\cite{bagchi1}
from the analyses of the total fraction of contacts between methyl groups on composition
and of the average number of methyl groups around a given group (see figures 9 and 10
in reference~\cite{bagchi1}). However, these authors claimed to observe a discontinuous 
transition termed as weak transition between two regimes, in contrast to our observations. 
A similar curve coming from the calculations for P2-TIP4P-2005 is given in figure~\ref{fig_8}~b as well.
In this case, there is a very weakly pronounced change of the inclination of the curve
between $X_d$ = 0.1667 and 0.2. However, it can hardly be interpreted as a
pronounced discontinuity.

We would like to finish this part by recalling that the predictions of the
OPLS-TIP4P-2005 model for density, for the excess mixing volume and related
partials as well as for energetic aspects of mixing of species are not satisfactory.
Therefore, it does not seem consistent to discuss anomalous behaviors by using 
solely certain aspects of microscopic structure from a particular model force field
that leads to a quite inaccurate thermodynamic and other properties.   
Evidently, our comments concerning the microscopic structure are not comprehensive.
It would be desirable to validate the structure by comparison with the 
experimental predictions for the structure factors in the entire composition range.
Unfortunately, as it was mentioned at the beginning of this subsection,
such data are not available in the literature, up to our best knowledge.

In contrast to microscopic structure, there are different possibilities to validate
the force field models against the experimental data. One of the possibilities is to discuss the
dynamic properties. Most common, computer simulations predictions are confronted 
with the experimental results for self-diffusion coefficients of species upon
the composition of the solutions in question.

\subsection{Self-diffusion coefficients of DMSO and water molecules}

\begin{figure}[h]
\begin{center}
\includegraphics[width=6.5cm,clip]{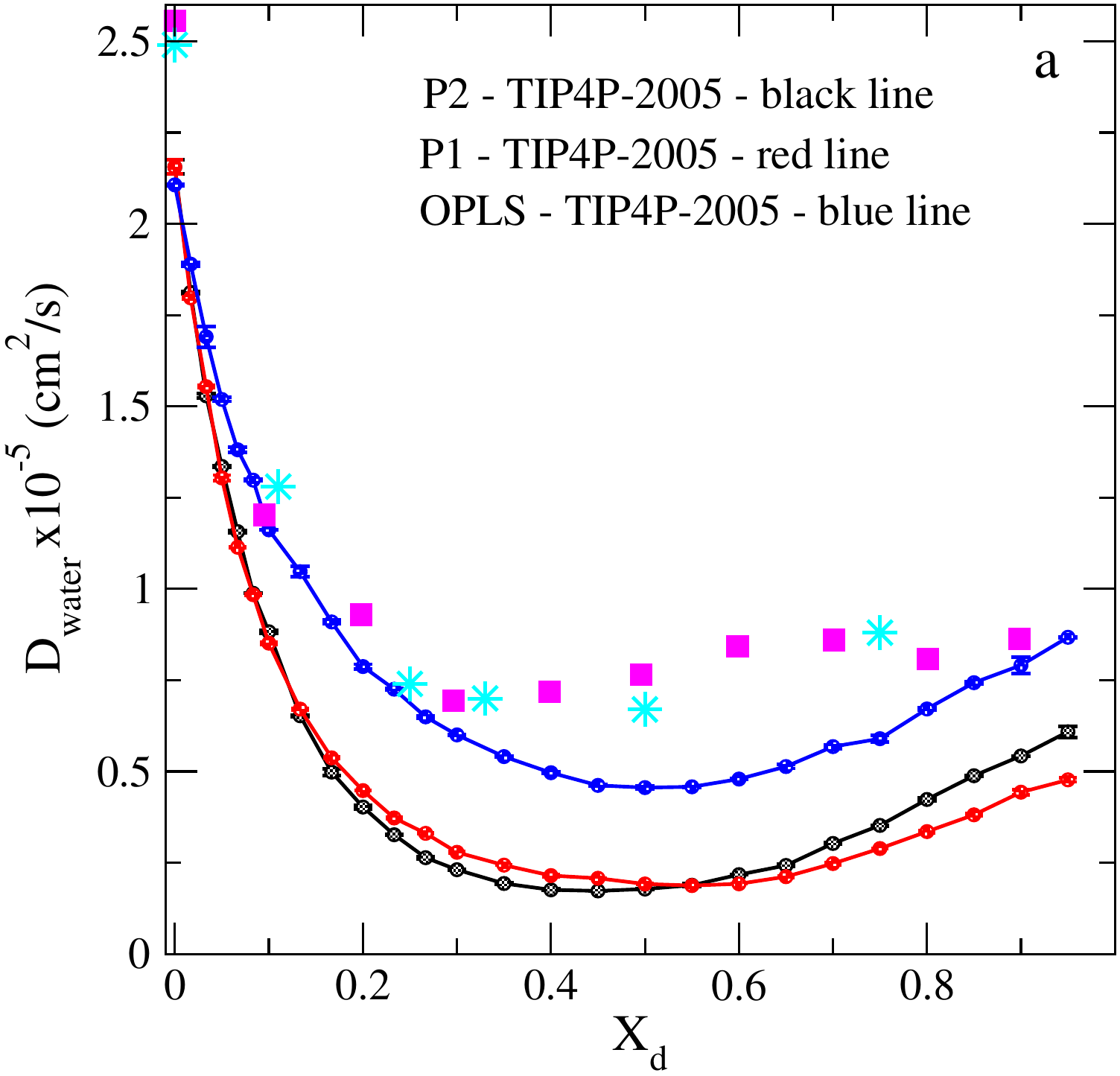}
\includegraphics[width=6.5cm,clip]{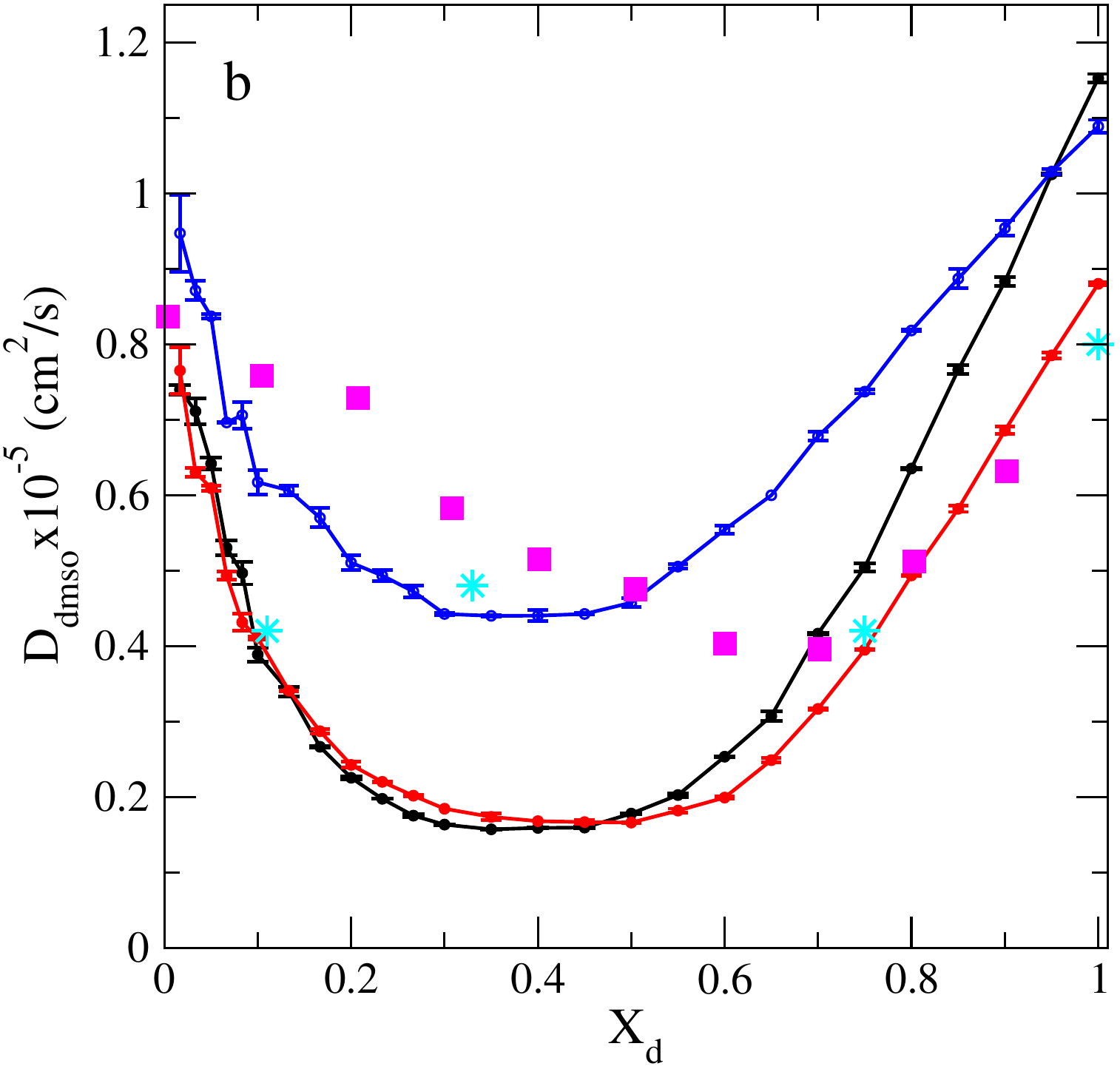}
\end{center}
\caption{(Colour online) Panels a and b: composition dependence of 
the self-diffusion coefficient  of water-DMSO mixtures
at  298.15~K for different combinations of DMSO and water models.
Experimental data are from reference~\cite{packer} (squares) and 
from reference~\cite{bordallo} (stars).
}
\label{fig_9}
\protect
\end{figure}

We proceed now to the description of the behavior of the self-diffusion 
coefficients of species, $D_i$ ($i = w$, DMSO), upon changing composition of the mixture.
In figure~\ref{fig_9}  we show the results obtained via  the Einstein relation,
\begin{equation}
D_i =\frac{1}{6} \lim_{t \rightarrow \infty} \frac{\rd}{\rd t} \vert {\bf r}_i(\tau+t)-{\bf r}_i(\tau)\vert ^2,
\end{equation}
where  $\tau$ denotes the time origin. Default settings of GROMACS were used for the separation of
the time origins and for the fitting interval. The experimental data were taken from 
references~\cite{packer,bordallo}. According to the experiments, the self-diffusion coefficient 
of water species decreases in magnitude starting from pure water value (at $X_d$ = 0) and reaches
minimum at $X_d \approx 0.3$. Next, for higher $X_d$, $D_w$ does not change much, but a certain
pecuiliarity is seen in the interval between 60\% and 80\%. Apparently, the behavior of 
$D_i(X_d)$ is determined by the evolution of density of the mixture and by the hydrogen bonding 
between species. Theoretical predictions for $D_w$ lead to a bit inaccurate, 
but still satisfactory agreement with experimental trends for water-rich mixtures. 
The OPLS-TIP4P-2005 model
looks the best, because it underestimates the density in the entire composition interval
starting from water-rich and ending up at DMSO-rich mixtures, cf. figure~\ref{fig_1}. 
This model yields a good value for $D_w$ close to $X_d= 1$, as well. Both, the P1-TIP4P-2005
and P2-TIP4P-2005 model yield qualitatively correct trends but essentially
underestimate $D_w$ in a wide interval of compositions.
Besides two factors mentioned just above, we can attribute such low values for $D_w$ to
the pronounced inaccuracy of description of the energetic factors of mixing of species
in the framework of these two models, cf. figure~\ref{fig_4}~b.
Better performance of the OPLS-TIP4P-2005 model for the excess mixing enthalpy 
apparently contributes to better description of~$D_w$.

According to the experimental data by Packer and Tomlinson~\cite{packer}, the
self-diffusion coefficient  of DMSO species has a minimum at $X_d \approx$ 0.7.
Besides, there is a peculiarity of $D_{\rm{DMSO}}$ in the interval of compositions corresponding
to water-rich mixtures. The model of the present study yields satisfactory predictions
for $D_{\rm{DMSO}}$ close to $X_d \approx 0$. Each of the models in question exhibits a weakly
pronounced peculiarity around $X_d \approx 0.1$. However, this peculiarity is much weaker
compared to what one observes from experimental points. On the other hand, close to $X_d \approx 1$,
the P1-TIP4P-2005 model describes $D_{\rm{DMSO}}$ pretty well. However, the minimum
predicted by the models occurs at intermediate compositions around 0.5, rather than at 0.7
from the experimental points.

To summarize, the models used in simulation perform better at ``wings'', i.e., close to
$X_d \approx 0$ and $X_d \approx 1$, rather than at intermediate compositions. Still,
the dynamic aspects of mixing are apparently better described for water-rich mixtures.
We have not performed a detailed investigation of the self-diffusion coefficient of
species for DMSO models combined with TIP4P/$\varepsilon$ model. It is known that the
diffusion coefficient of pure water in the framework of TIP4P-2005 and TIP4P/$\varepsilon$ model
is very similar~\cite{alejandre}. Thus, we do not expect to obtain an improvement of
the observed trends by changes of the water models of this type. 

The missing elements worth to explore are various. Namely, it is necessary to confirm
the present results by using alternative calculations of the self-diffusion coefficients
via the velocity auto-correlation functions. In addition, one should attempt to calculate the 
relaxation times and possibly the power spectra, similarly to e.g.,~\cite{galicia2}, because of the
availability of experimental data. It would provide ampler insights into the
adequacy of the force fields.


\subsection{Static dielectric constant of water-DMSO mixtures}

Now, proceed to the results for the static dielectric constant. 
It is one of the most difficult properties to deal with. Recently,
we explored the dielectric constant for water-DMSO mixtures by using the \mbox{SPC-E}~water 
model~\cite{gujt}. Therefore, the present discussion will be brief and
solely novel findings will be underlined.

The long-range, asymptotic behavior of correlations between the molecules 
possessing a dipole moment is described by the dielectric constant, $\varepsilon$,
which is calculated from the time-average of the fluctuations of the total
dipole moment of the system~\cite{martin} as follows,
\begin{equation}
\varepsilon=1+\frac{4\piup}{3 k_{\rm{B}} TV} \left(\langle \bf M^2 \rangle -\langle \bf M \rangle^2 \right),
\end{equation}
where $k_{\rm{B}}$ is the Boltzmann constant and $V$ is the simulation cell volume. 
In this work we took
care that each of the runs is of sufficient time extension (more than 100~ns in all cases), in order
to obtain reasonable statistics. The final value was recorded. In contrast
to reference~\cite{gujt}, the averaging over blocks was not performed. 
Besides the stability of the dielectric constant value along the run,
we verified  the temporal behavior of the average total polarization of the system 
to ensure that each of the components ($x$, $y$ and $z$) is close to zero.
Evidently, the number of molecules, type of thermostat and
barostat, precision of the long-range interactions summation contribute into the
accuracy of the evaluation of the dielectric constant.

\begin{figure}[h]
	\begin{center}
		\includegraphics[width=6.5cm,clip]{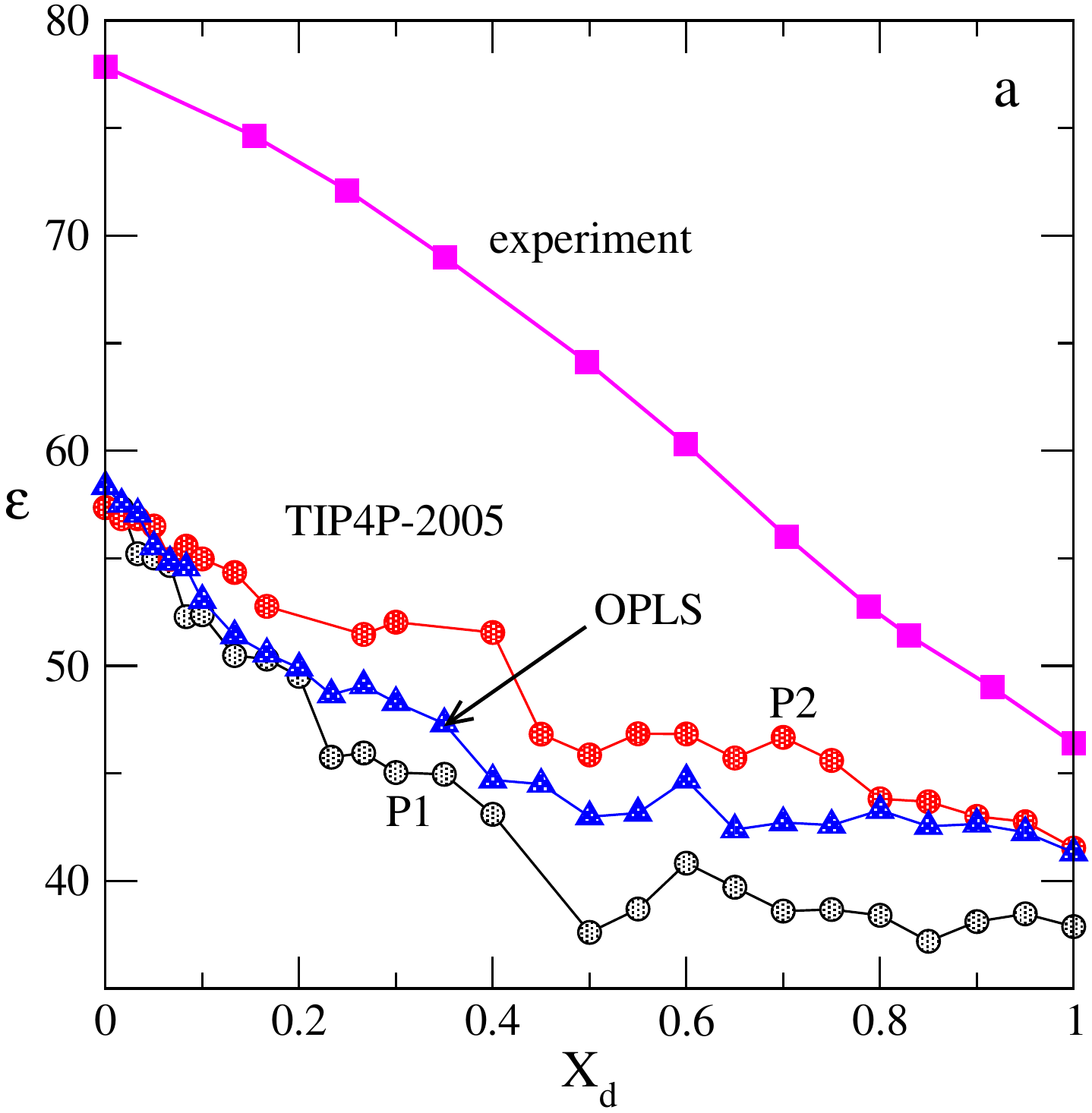}
		\includegraphics[width=6.5cm,clip]{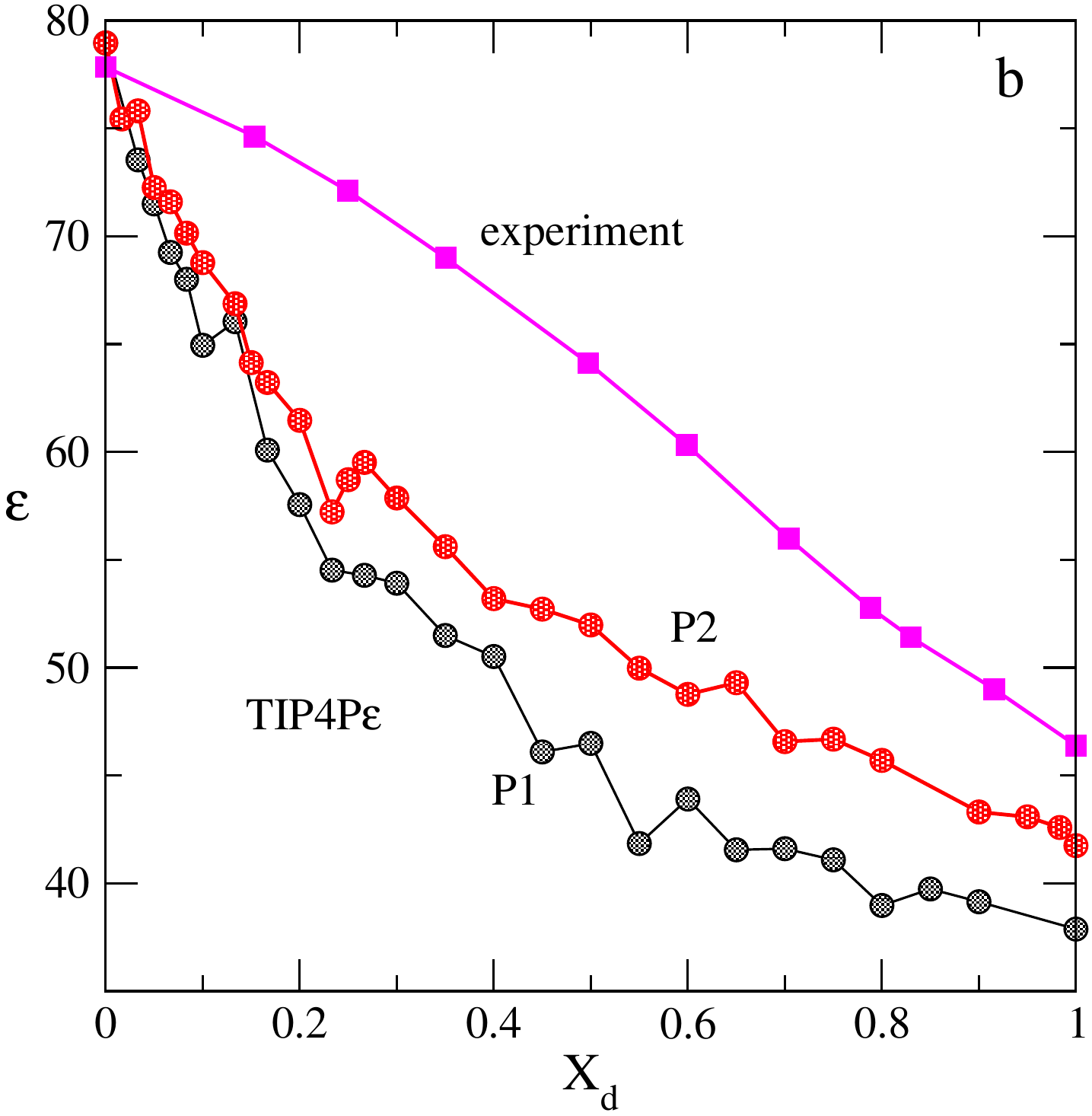}
	\end{center}
	\caption{(Colour online) Panels a and b: composition dependence of
		the static dielectric constant  of water-DMSO mixtures
		at  298.15~K for different combinations of DMSO and water models.
		Experimental data are from reference~\cite{plowas}.
	}
	\label{fig_10}
	\protect
\end{figure}

The experimental data are taken from reference~\cite{plowas}. 
The dielectric constant decreases from a high value for water to a 
much lower value for pure DMSO. According to the
experiment, the inclination of the curve for $\varepsilon(X_d)$ changes around
$X_d \approx 0.2$. Moreover, the dielectric constant is almost linear for $X_d > 0.8$.
It is known that the TIP4P-2005 does not provide a reasonable description of
the dielectric constant for pure water. Thus, if one changes the DMSO model,
the results do not improve on the water-rich side of the composition, panel~a of figure~\ref{fig_10}.
On the DMSO-rich side, the OPLS and P2 models, if combined with TIP4P-2005
water model, yield not very bad values for $\varepsilon$. However, general trends
of decay of $\varepsilon$ with increasing $X_d$ are not satisfactory. In spite
of long time of simulations, the values for $\varepsilon$ strongly fluctuate
along $X_d$ axis, such that it is impossible to capture any confiable peculiarities.
Substantial improvement of the results is reached, if one substitutes TIP4P-2005
water model by the TIP4P/$\varepsilon$ model. Then, the values for dielectric
constant essentially improve for water-rich mixtures. Moreover, the simulation
data fluctuate less upon changing the composition $X_d$. Two regions of
behavior can be identified. Namely, the simulation data predict a fast decay
of $\varepsilon$ for water-rich mixtures and a slower decay in the opposite interval
of compositions, say at $X_d > 0.3$. The experimental points, by contrast, show a
slower decay for a water-rich mixture and a slightly weaker decay at higher values of $X_d$.
After long runs (150~ns), we obtained with confidence a peculiarity in
the values of $\varepsilon$ in the interval of $X_d$ between 0.2 and 0.3. It is
similar by shape to the behavior of mean square deviation of total dipole moment
on composition plotted in figure~3 of reference~\cite{bagchi2} at $X_d$ between 0.1 and 0.15
with another force field. These trends, however, result from the modelling rather
than from the experimental observations.
In view of the present findings, the interpretation of data in terms of structural phase transition 
in laboratory systems seems to be questionable.

Apparently, the inaccuracy of the description of energetic trends of mixing
the water and DMSO species in the framework of P1 and P2 DMSO models leads to
unsatisfactory dependence of the dielectric constant on $X_d$. In qualitative terms,
the conclusion is similar to what was mentioned above concerning the self-diffusion
coefficients of species. However, it is worth to complement the calculations of
the dielectric constant by the exploration of the behavior of the Kirkwood factor. 
There is room for intent improvement of the data. One can consider larger systems,
exclude the fluctuations of pressure by switching to the  canonical, NVT, ensemble,
take larger values for cut-off distance. Application of these technical issues
does not guarantee better results. Seemingly, present non-polarizable DMSO models 
with the present schemes of taking into account the long-range electrostatic interactions
are incapable of reproducing the experimental data. Definitely, the situation improves if
one applies even simple  polarizable DMSO model, see e.g.~\cite{bachmann}. Then, one
would expect that the combination of the TIP4P/$\varepsilon$ water with even
simplified polarizable DMSO model would be profitable. These issues guarantee future studies.

\begin{figure}[h]
	\begin{center}
		\includegraphics[width=6.5cm,clip]{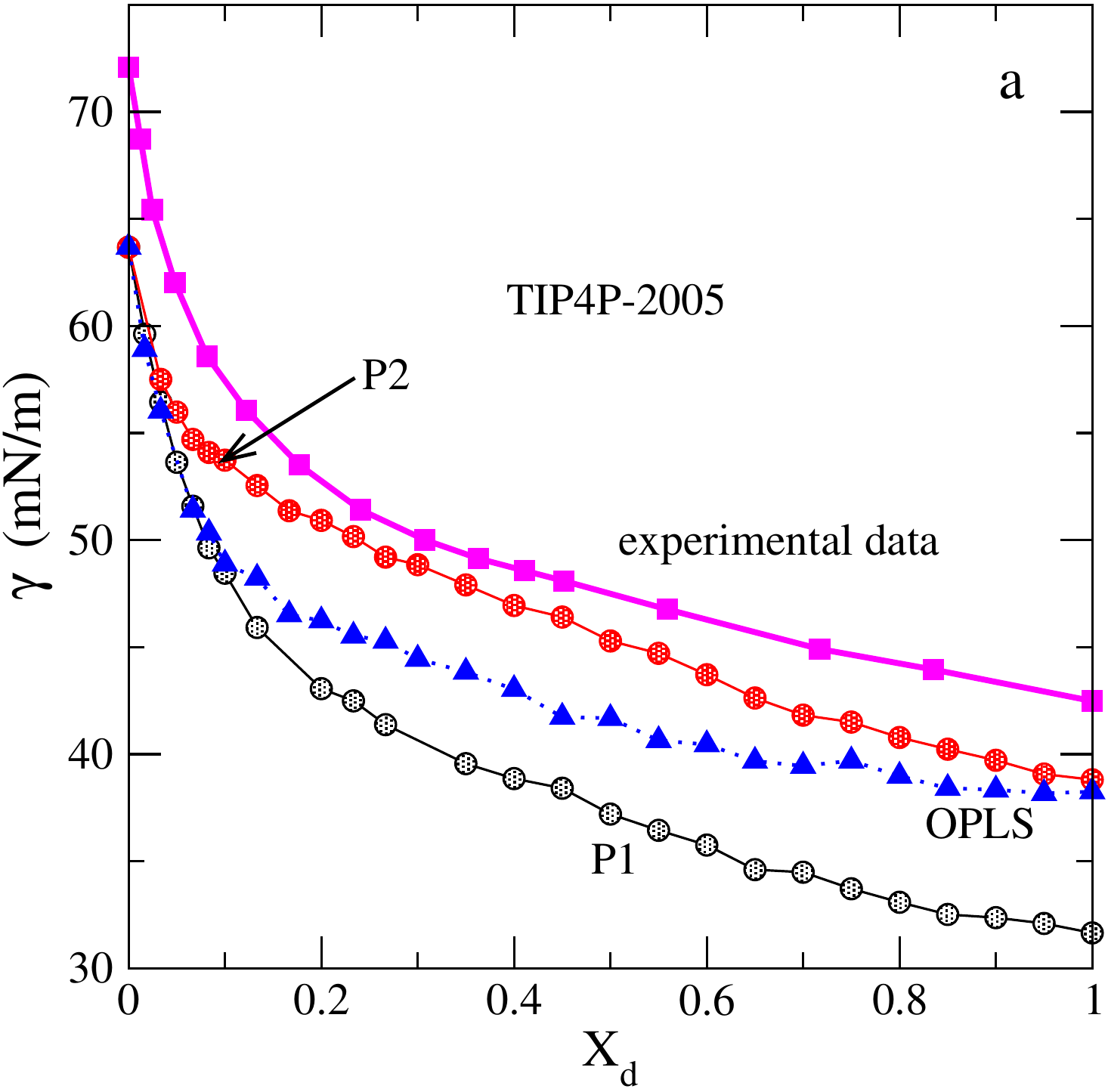}
		\includegraphics[width=6.5cm,clip]{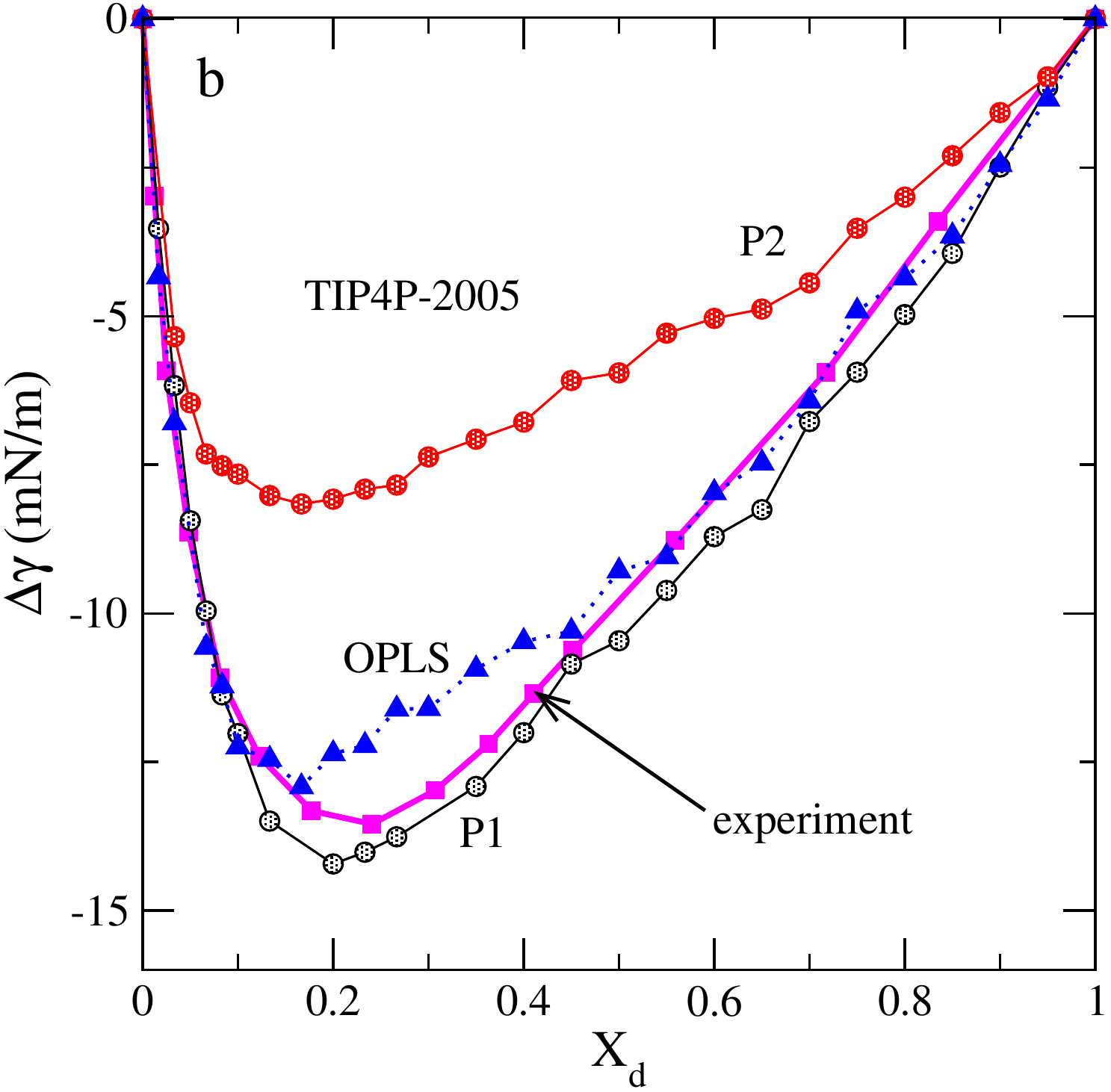}
	\end{center}
	\caption{(Colour online) Composition dependence of the surface tension (panel a) and of the excess
		surface tension (panel b) for different combinations of DMSO and water models.
		Experimental data are taken from reference~\cite{fazli}.
	}
	\label{fig_11}
	\protect
\end{figure}

\subsection{Surface tension of water-DMSO mixtures on composition}

Our final remarks concern the behavior of the surface tension of water-DMSO 
mixtures. The  surface tension calculations at each
composition were performed by taking the final configuration of particles 
from the isobaric run. Next, the box edge along $z$-axis was extended by a factor of 3, 
generating a rectangular box with a liquid slab
and two liquid-mixture-vacuum interfaces in the $x-y$ plane, 
in close similarity to the procedure applied in~\cite{vanderspoel}.
The total number of molecules is sufficient to yield sufficiently big area of the
$x-y$ face of the liquid slab. The elongation of the liquid slab along
$z$-axis is satisfactory as well.
The executable file was modified by deleting
a fixed pressure condition preserving the V-rescale thermostatting with the same parameters
as in the NPT runs. Other corrections were not employed.
The values for the surface tension, $\gamma$, follow from the combination of the time
averages for the components of the pressure tensor,
\begin{equation}
\gamma = \frac {1}{2} L_z \left\langle{\left[ P_{zz}-\frac{1}{2} \left( P_{xx}+P_{yy} \right) \right]}\right\rangle,
\end{equation}
where $P_{ij}$ are the components of the pressure tensor along $i,j$ axes, and  $\langle...\rangle$
denotes the time average.
We performed five runs at a constant volume, each piece of 10~ns, and
obtained the result for $\gamma$ by taking the block average. 
The experimental results were taken from reference~\cite{fazli}.
The data show that the surface tension rapidly decreases from the pure water
value at $X_d=0$ till $X_d \approx 0.25$. Next, at higher values of $X_d$, the values
for $\gamma$ decrease more slowly, the curve behaves almost linearly in that interval
of compositions, panel~a of figure~\ref{fig_11}. The excess surface tension from experiment
exhibits minimum at $X_d \approx 0.25$. All the models in question behave similarly 
for water-rich mixtures. The decay of $\gamma$ is better reproduced by P1-TIP4P-2005
and OPLS-TIP4P-2005 models, rather than by P2-TIP4P-2005 model. The absolute values for
$\gamma$ are underestimated, but can be improved by including the long-range corrections. 

Concerning the behavior of $\gamma$ for DMSO-rich mixtures, one should note that
the P2-TIP4P-2005 model leads to the results most close to the experimental data.
However, the excess surface tension is underestimated by this model intermediate
compositions.
On the other hand, the P1-TIP4P-2005 and OPLS-TIP4P-2005 reproduce the excess surface 
tension rather well in the entire interval of compositions, panel b of figure~\ref{fig_11}.
The shape of the surface tension on $X_d$ is quite well reproduced by these two
models, just the absolute values of $\gamma$ are underestimated. Thus, a
reasonable description of $\gamma$ and its excess simultaneously by the models of this
study restricts to water-rich mixtures.
Still, it seems that a better agreement with experimental data may result from the
changes of fine details of simulation procedure as documented in reference~\cite{labastida}.

\section{Summary and conclusions}

Recently we studied the properties of water-DMSO liquid mixtures in the entire 
composition interval by molecular dynamics computer simulations of the DMSO models
combined with the SPC-E water model~\cite{gujt}. In the present work, we explore the same systems
with the same methodology but using the non-polarizable, P1, P2 and OPLS united atom DMSO models
in conjunction with the TIP4P-2005 and TIP4P/$\varepsilon$ water models. Concerning
the DMSO models, we observed that the flexibility of angles, keeping the bond
constrained, requires to include the dihedral angle as recommended in reference~\cite{oostenbrink}.
The systems in question are studied at room temperature, 298.15~K and atmospheric
pressure, 0.1~MPa, as previously. 

A set of novel findings was obtained and discussed in every detail. Namely,
in the present work we obtained the excess partial molar volumes and apparent molar
volumes. Analyses of these properties permitted us to capture an anomalous behavior
observed at low DMSO molar fraction in agreement with experimental results. 
Besides, the excess partial molar enthalpy of each species was obtained and 
compared with experimental findings. Particular emphasis was put on the 
elucidation of the behavior of various radial distribution functions in water-rich 
mixtures. The coordination numbers and the fractions of molecules participating in 
hydrogen bonds were involved in the interpretation of the microscopic structure.
The self-diffusion coefficients of species were obtained and compared with 
experimental data. Next, the static dielectric constant was calculated and
compared with experimental results. Our final focus was on elucidating the behavior of 
the surface tension on the composition of water-DMSO mixtures and of the excess
surface tension. Their trends were compared with experimental data.
A very detailed validation of the predictions of 
the properties resulting from a set of models for water-DMSO  solutions 
permit to make conclusions w.r.t. their applicability.

Among the variety of our results, solely the discussion of changes of the
microscopic structure lacks comparison with experiments. In order to do
that, one needs to have data concerning the total and desirably the 
partial structure factors in the entire composition interval. 
Then, it would be possible to perform the analyses
similar to what was done in this laboratory for water-methanol mixtures~\cite{galicia2}.
We are not aware of such type of  results for water-DMSO mixtures at present.
This part of study of water-DMSO mixtures  might benefit from the reverse Monte Carlo modelling
in order to elucidate fine features of the hydrogen bonded network, see  e.g.,~\cite{pusztai1}.
An illuminating finding about the existence and lifetime
of frequently discussed complexes or associates may be expected.

According to our findings, one can distinguish two types of properties. 
Namely, for one type, the role of fluctuations is less important.
Evidently,  at a fixed pressure the systems' volume fluctuates, 
but density as an average over a set of computer simulation runs is well defined.  
The density dependence on composition compares well with experiment.
Consequently, one is able to describe well the excess partial molar volumes
and apparent molar volumes and definitely prove the existence of an anomalous
behavior for water-rich water-DMSO mixtures. Previously, this kind
of behavior was obtained and discussed for water-methanol 
mixtures~\cite{mario,chechko}.

On the other hand, under the same conditions to obtain the isothermal compressibility,
for example, requires time average of fluctuations. 
Consequently, this property is more difficult to obtain.  
Similarly, a disappointing situation is observed in the calculations of the
static dielectric constant. It results from the time evolution of the total 
dipole moment fluctuations. If $\varepsilon$ is calculated from the NPT
runs, the volume fluctuations contribute to the inaccuracy of the results. 
In addition, at a constant density, the values for $\varepsilon$ are affected
by the choice of the number of particles in the simulation box, by the
assumed cut-off of the interactions and by the scheme of summation of long-range
interaction effects. 

One should have these difficulties in mind while interpreting the changes of various
properties on composition. Specifically, we observed that peculiarities of the
behavior of different descriptors hardly permit to interpret them as discontinuous
structural transitions in laboratory systems, see e.g., figures~3 and 5 of reference~\cite{bagchi2} 
and figures~8 and 9 of reference~\cite{bagchi1}, unless a precise description
of the underlying properties in terms of microscopic structure and/or the
static dielectric constant is obtained by computer simulations.

To summarize, we would like to mention that each of the explored models
was designed, or say parametrized, using a restricted number of different criteria at certain
thermodynamic conditions. Thus, it is difficult to expect that modelling
can be successful in a wide range of temperature and pressure. Similarly,
we observed ``non-universal'' performance of the models along the composition axes.
It appears that the P2-TIP4P-2005 model provides a reasonable description
of a wide set of properties for water-rich mixtures, except the static dielectric
constant. However, this behavior can be mitigated by using the P2 - TIP4P/$\varepsilon$ model.
In addition, we would like to note that the P1-TIP4P-2005 model is better
than the P2-TIP4P-2005 model for some properties of DMSO-rich mixtures and 
describes the surface tension and the excess surface tension reasonably well.
Hence, one should choose the working model dependent on how concentrated 
DMSO solution is necessary to deal with, besides the pressure and temperature values
of the given setup.
The OPLS-TIP4P-2005 model is successful solely for a few properties.
Hence,  possible modifications of the parameters and combination rules should be attempted.

Finally, we would like to enlist a few missing elements to extend our 
knowledge of the properties of water-DMSO mixtures.  
It would be profitable to explore various velocity
auto-correlation functions to enhance the understanding of dynamic properties, as well as to
investigate Kirkwood factors and possibly the frequency dependent dielectric constant
to understand the dielectric properties better.
Another missing issue is to explore the  cooperativity of hydrogen bond network
in terms of certain complex elements, see e.g.,~\cite{pusztai2}, and hydrogen bonds life times.

\section*{Acknowledgements} 
O.~P. acknowledges helpful discussions with Prof. M. Holovko
and Dr. T. Patsahan concerning the interpretation of the dielectric constant results
and some technical aspects of simulations, respectively.


\ukrainianpart

\title{ Ще раз про концентраційні залежності властивостей рідких сумішей вода-диметилсульфоксид. Моделювання методом молекулярної динаміки. }
\author{M. Агілар\refaddr{label1}, 
	Е. Домінгес\refaddr{label2},
	O. Пізіо\refaddr{label1}
}

\addresses{
	\addr{label1}Інститут хімії, Національний автономний університет Мехіко,
	Circuito Exterior, 04510, Мехіко, Мексика
	\addr{label2} Інститут дослідження матеріалів, Національний автономний університет Мехіко,
	Circuito Exterior, 04510, Мехіко, Мексика}

\makeukrtitle
\begin{abstract}
У цій роботі ми повертаємося до розгляду концентраційних залежностей головних властивостей рідких сумішей вода-ДМСО
з використанням комп'ютерного моделювання методом ізобарично-ізотермічної молекулярної динаміки. 
Розглянуто набір неполяризовних напівгнучких моделей молекул ДМСО
у поєднанні з моделями води TIP4P-2005 і TIP4P/$\varepsilon$. Обчислення проведено при атмосферному тиску 0.1013 MПа та кімнатній температурі 298.15~K. Наведено концентраційні залежності густини, надлишкового об'єму змішування, надлишкової ентальпії змішування, парціальних молярних об'ємів і парціальної молярної ентальпії складників суміші,
видимих молярних об'ємів. Також досліджено концентраційні залежності самодифузії складників, статичної діелектричної сталої та поверхневого натягу. Зміни мікроскопічної структури суміші в залежності від її складу проаналізовано на основі радіальних функції розподілу, координаційних чисел та часток молекул, пов'язаних водневими зв'язками.
Метою є зафіксувати особливості змішування складників суміші при зміні молярної частки ДМСО, а також можливі аномалії досліджених характеристик. Проведено оцінку придатності різних комбінацій моделей складників суміші для того, щоб з'ясувати необхідність у подальшому вдосконаленні.
	
\keywords{ молекулярна динаміка, суміші вода-диметилсульфоксид (ДМСО), поверхневий натяг, діелектрична стала, парціальні молярні об'єми, парціальні молярні ентальпії }
	
\end{abstract}

\lastpage  
\end{document}